\documentclass[reprint,aps,twocolumn,showpacs,superscriptaddress,groupedaddress,nofootinbib]{revtex4-2} 
\usepackage{graphicx}  % needed for figures
\usepackage{dcolumn}   % needed for some tables
\usepackage{bm}        % for math
\usepackage{amssymb}   % for math
\usepackage{amsmath}   % for math
\usepackage{multirow}
\usepackage{hyperref}
\usepackage{braket}
\usepackage{subfigure}
\usepackage{array}
\usepackage[math]{cellspace}
\usepackage[utf8]{inputenc}
\usepackage{cleveref}
\usepackage{algorithm}
\usepackage{algpseudocode}
\algrenewcommand\algorithmicrequire{\textbf{Input:}}
\algrenewcommand\algorithmicensure{\textbf{Output:}}

\usepackage{color}
\usepackage[normalem]{ulem}

\usepackage{url}
\hypersetup{
colorlinks=true,
linkcolor=magenta,
urlcolor=magenta,
citecolor=blue
} 
\begin{document}
\hspace{5.2in} %\mbox{IFIC/21-22}

\title{Learning Trivializing Flows}

\date{\today}

\author{D. Albandea\footnote{Hi}}
\email{david.albandea@uv.es}
\affiliation{IFIC (CSIC-UVEG), Edificio Institutos Investigaci\'on, 
Apt.\ 22085, E-46071 Valencia, Spain}
\author{L. Del Debbio}
\affiliation{Higgs Centre for Theoretical Physics, School of Physics and
Astronomy, The University of Edinburgh, Edinburgh EH9 3FD, UK}
\author{P. Hern\'andez}
\affiliation{IFIC (CSIC-UVEG), Edificio Institutos Investigaci\'on, 
Apt.\ 22085, E-46071 Valencia, Spain}
\author{R. Kenway}
\affiliation{Higgs Centre for Theoretical Physics, School of Physics and
Astronomy, The University of Edinburgh, Edinburgh EH9 3FD, UK}
\author{J. Marsh Rossney}
\affiliation{Higgs Centre for Theoretical Physics, School of Physics and
Astronomy, The University of Edinburgh, Edinburgh EH9 3FD, UK}
\author{A. Ramos}
\affiliation{IFIC (CSIC-UVEG), Edificio Institutos Investigaci\'on, 
Apt.\ 22085, E-46071 Valencia, Spain}

\begin{abstract}
The recent introduction of Machine Learning techniques, especially Normalizing Flows, for the sampling of lattice gauge theories has shed some hope on improving the sampling efficiency of the traditional Hybrid Monte Carlo (HMC) algorithm. In this work we study a modified HMC algorithm that draws on the seminal work on trivializing flows by L\"uscher. Autocorrelations are reduced by sampling from a simpler action that is related to the original action by an invertible mapping realised through Normalizing Flows models with a minimal set of training parameters. We test the algorithm in a $\phi^{4}$ theory in 2D where we observe reduced autocorrelation times compared with HMC, and demonstrate that the training can be done at small unphysical volumes and used in physical conditions.  We also study the scaling of the algorithm towards the continuum limit under various assumptions  on the network architecture.
\end{abstract}
\maketitle
% \tableofcontents

\section{Introduction}
\label{sec:intro}

%Lattice field theories (LFTs) suffer from critical slowing down (CSD) when
%approaching a critical point of the theory, where the correlation length of the
%ystem diverges. 
Lattice Field Theory admits a numerical approach to the study of
non-perturbative properties of many field theories by using Markov Chain Monte
Carlo (MCMC) techniques to generate representative samples of field
configurations and computing expectation values.  However, standard MCMC
algorithms suffer from a phenomenon known as \textit{critical slowing down},
whereby the autocorrelation time of the simulation increases dramatically as the
continuum limit is approached.  In many theories of interest, including Quantum
Chromodynamics (QCD), this problem is exacerbated by the effect of
\emph{topology freezing}
\cite{Campostrini1992,Vicari1993,DelDebbio:2004xh,Schaefer2011,Flynn2015,Bonati2018a,DelDebbio2002,Alles1996,Luscher:2010iy,Virotta2010}.
Autocorrelation times of topological observables have been shown to scale
exponentially with the inverse lattice spacing, $a^{-1}$, for the CP$^{N-1}$
model \cite{DelDebbio:2004xh}, and at least polynomially with $a^{-6}$ for
lattice QCD \cite{Luscher:2010iy}. 

\textit{Trivializing maps} are invertible field transformations that map a
complicated theory to a trivial one, i.e. to a limit in which the field
variables decouple and sampling is extremely efficient.  L\"uscher
\cite{LuscherTrivializingMaps} originally proposed an augmentation of the Hybrid
Monte Carlo (HMC) algorithm in which an
approximate trivializing map is used to reduce autocorrelation times.  However,
when tested against CP\(^{N-1}\) models, it was reported that the quality of
this approximation, which involved computing the first few terms of a power
series, was not sufficient to improve the scaling of the computational cost
towards the continuum limit with respect to standard HMC \cite{Schaefer2011}.

The recent introduction of Machine Learning techniques for the sampling of
lattice field theories has opened a new avenue to address critical slowing down
in lattice field theories
\cite{Albergo2019,Kanwar2020,Nicoli2020,Boyda2020,Albergo2021,Albergo2022,Abbott2022}.
A class of Machine Learning models known as Normalizing Flows are also
invertible transformations that are parametrised by neural networks (NNs) and
can hence be `trained' to approximate a desired mapping
\cite{Tabak2010,Tabak2012,Rezende2015}.  Albergo, Kanwar and Shanahan
\cite{Albergo2019} first demonstrated that direct sampling from a well-trained
Normalizing Flow, combined with some form of reweighting such as a Metropolis
test, produces unbiased samples of field configurations while completely
avoiding critical slowing down.  However, experiments with simple architectures
have suggested that the overhead cost of training models to achieve a fixed
autocorrelation time scales extremely poorly towards the continuum limit
\cite{DelDebbio:2021qwf}. 

In this work we investigate an algorithm inspired by the original idea of
L\"uscher, but where a Normalizing Flow is used to approximate the trivializing
map.  Given the high training costs associated with the direct sampling
strategy, we pose the question: is it possible to improve the scaling of
autocorrelation times in HMC using minimal models that are cheap to train?  To
answer this we benchmark our method against standard HMC on a two-dimensional
\(\phi^{4}\) model.

The paper is organised as follows: in Section~\ref{sec:org9d8cd18} we briefly
review trivializing flows before describing the algorithm that is the focus of
this work; in Section~\ref{sec:model-and-setup} we describe the experimental setup,
which includes details about the $\phi^4$ theory, the Normalizing Flow
architectures, and the HMC component of the algorithm; in
Section~\ref{sec:orgbc47492} we provide the results of our experiments and compare
the computational cost scaling against standard HMC. This work is based on
results previously reported in Reference~\cite{Albandea:2022fky}.
\newpage

\section{Learning trivializing flows}
\label{sec:org9d8cd18}

\subsection{Trivializing flows}
\label{sec:triv-flows}
Consider the expectation value of an observable in the path integral formalism
of a quantum field theory in Euclidean spacetime,
\begin{align}
\label{eq:Zint}
\langle \mathcal{O} \rangle = \frac{1}{\mathcal{Z}} \int_{ }^{ } \mathcal{D}\phi
\;  \mathcal{O}(\phi) \, e^{-S(\phi)},
\end{align}
where \(\mathcal{O}(\phi)\) is an observable defined for the field configuration
\(\phi\), \(S(\phi)\) is the action of the theory, $\mathcal{Z}$ is its
partition function,
\begin{align}
    \mathcal{Z} = \int \mathcal{D}\phi\; e^{-S(\phi)},
\end{align}
and \(\mathcal{D}\phi\) is the integration measure,
\begin{align}
\mathcal{D}\phi = \prod_{x} d\phi_{x}.
\end{align}
The probability of a field configuration $\phi$ is given by the 
Boltzmann factor 
\begin{equation}
    \label{eq:TargetProb}
    p(\phi) = \frac{1}{\mathcal{Z}}\, e^{-S(\phi)}\, .
\end{equation}
We will refer to $p$ as the {\em target distribution}. 
A change of variables \(\tilde{\phi}=\mathcal{F}^{-1}(\phi)\)
in Equation~(\ref{eq:Zint}) yields
\begin{align}
\langle \mathcal{O} \rangle = \frac{1}{\mathcal{Z}} \int_{ }^{ } D \tilde{\phi
}\; \mathcal{O}(\mathcal{F}(\tilde{\phi})) e^{-S[\mathcal{F}(\tilde{\phi })] + \log \det J[\mathcal{F}(
\tilde{\phi } )]},
\end{align}
where $J[ \mathcal{F}(\tilde{\phi}) ]$ is the Jacobian coming from the change in
the integral measure, $\mathcal{D}\phi=\mathcal{D}\tilde{\phi}\,\det J[ \mathcal{F} ]$.
If \(\mathcal{F}\) is chosen such that the effective action for the transformed field,
\begin{align}
\label{eq:complete-triv-map}
\tilde{S}(\tilde{\phi}) \equiv S[\mathcal{F}(\tilde{\phi})] - \log \det J[\mathcal{F}(\tilde{\phi})] \, ,
\end{align}
describes a non-interacting theory, then \(\mathcal{F}\) is known as a trivializing map.

In Reference~\cite{LuscherTrivializingMaps} trivializing maps for gauge theories were constructed as flows
\begin{align}
\label{eq:triv-flow-eq}
\dot{\phi}_t \equiv T[t, \phi_t],
\end{align}
with boundary condition
\begin{align}
\phi = \phi_0.
\end{align}
The trivializing map is defined as
\begin{align}
\tilde{\phi} = \mathcal{F}^{-1}(\phi) = \phi_{1}.
\end{align}
Though not known in closed form, the kernel \(T\) of the trivializing flow can be expressed as 
a power series in the flow time \(t\).  In practice, this power series was 
truncated at
leading order and the flow integrated numerically, resulting in an 
approximate
trivializing map where the effective action in 
Equation~(\ref{eq:complete-triv-map}) is
still interacting in general. 
Nevertheless, $\tilde{S}$ ought to be easier to sample than $S$.

The algorithm introduced in Reference~\cite{LuscherTrivializingMaps} is essentially the HMC algorithm applied to the flowed field variables $\tilde{\phi}$.
This algorithm was tested for the CP$^{N-1}$ model, which suffers from
topology freezing \cite{DelDebbio:2004xh}, by Engel and Schaefer
\cite{Schaefer2011}.  The conclusion of the study was that, although there was a
small improvement in the proportionality factor, the overall scaling of the
computational cost towards the continuum did not change with respect to standard HMC.

\subsection{Flow HMC (FHMC)}
\label{sec:norm-flows}
Normalizing Flows are a machine learning sampling technique first applied to
lattice field theories in Reference~\cite{Albergo2019}. The idea is similar to that
of the trivializing map. Starting from an initial set of configurations
\(\{z_{i}\}_{i=1}^{N}\) drawn from a probability distribution where 
sampling is easy,\footnote{An example of such {\em easy} distribution 
could be a multi-dimensional normal distribution.} 
\begin{align}
    \label{eq:EasyProb}
    z_{i}\sim r(z),
\end{align}
a transformation $\phi = f^{-1}(z)$ is applied  so
that  the transformed configurations \(\{\phi_{i}\}_{i=1}^{N} \equiv
\{f^{-1}(z_{i})\}_{i=1}^{N}\) follow the new probability distribution
\begin{align}
    p_{f}(\phi) = r\bigl(f(\phi)\bigr)\,  
    \left| \det \frac{\partial f(\phi)}{\partial \phi} \right|.
\end{align}
The probability density $p_f$ is called the {\em model distribution}.
The transformation \(f\) is implemented via NNs with a set of
trainable parameters $\{\theta_{i}\}$ which have been optimised so that
\(p_{f}(\phi)\) is as close as possible to the target distribution
\(p(\phi) = e^{-S(\phi)} / \mathcal{Z}\), i.e. the distribution of the theory we
are interested in.  The determinant of this transformation can be easily
computed if the network architecture consists of coupling
layers~\cite{Dinh2014,Dinh2016,Kingma2018} with a checkerboard pattern, as
explained in Reference~\cite{Albergo2019}.  Normalizing Flows are therefore NN
parametrisations of trivializing maps. 

Ideally, the NNs would be trained such that the K\"ullbach-Leibler (KL)
divergence~\cite{Kullbach1951} between the model and the target distribution,
\begin{align}
    D_{\text{KL}}(p_{f} || p) = \int \mathcal{D}\phi \,
    p_{f}(\phi) \log \frac{p_{f}(\phi)}{p(\phi)},
\end{align}
is minimised.
The KL divergence is a statistical estimator satisfying \(D_{\text{KL}}(p_{f} || p) \ge 0\)
and 
\[
    D_{\text{KL}}(p_{f} || p) = 0 \iff p_{f} = p\, .
\]
However, the partition function \(\mathcal{Z}\) appearing in \(p(\phi)\) is generally not known, so in practice one minimises a shifted KL divergence,
\begin{align}
    \label{eq:LossFun}
    L(\theta) \equiv&\, D_{\text{KL}}(p_{f} \mid \mid p) - \log
\mathcal{Z} \nonumber \\
    =& \int \mathcal{D}\phi\, p_{f}(\phi) \left[ \log
p_{f}(\phi) + S(\phi) \right].
\end{align}
This loss function can be stochastically estimated by drawing samples \(\phi_{i}
\sim p_{f}(\phi)\) from the model; there is no requirement to have a set of
existing training data.
Since $f$ is differentiable, the loss can be minimised using standard gradient-based optimisation algorithms such as stochastic gradient descent and ADAM~\cite{Kingma2014}, the latter of which we used in this study.
The absolute minimum of the loss function occurs when \(L(\theta) = -\log\mathcal{Z}\), where \(p_{f} = p\).
In practice this minimum is unlikely to be achieved due to both the limited \textit{expressivity} of the model and a finite amount of training, but one expects to have an approximate trivializing map at the end of the training, i.e. \(p_{f} \approx p\).

The Normalizing Flow model generates configurations that are distributed according to $p_{f}$, not $p$.
To achieve unbiased sampling from $p$, the work in Reference~\cite{Albergo2019} embeds the model in a Metropolis--Hastings (MH) algorithm~\cite{Metropolis1953,Hastings1970}, where $p_{f}$ serves as the proposal distribution.
Since the proposals are statistically independent, the only source of autocorrelations are rejections.
However, training models to achieve a fixed low MH rejection rate can become prohibitively expensive for large systems and long correlation lengths~\cite{DelDebbio:2021qwf}.

In contrast, in this work we propose to use the trained model as an
approximation to the trivializing map in the implementation of the trivializing
flow algorithm described in Section~\ref{sec:triv-flows}. Thus we identify
\begin{align}
\mathcal{F} = f^{-1},
\end{align}
so that the expectation value in Equation~(\ref{eq:Zint}) becomes,
\begin{align}
    \label{eq:pathintegral-chvar}
    \langle \mathcal{O} \rangle =&\, \frac{1}{\mathcal{Z}} 
    \int_{ }^{ } D \tilde{\phi }\; \mathcal{O}\left(f^{-1}(\tilde\phi)\right)\, 
    e^{-S[f^{-1}(\tilde{\phi })] + \log \det J[f^{-1}(
    \tilde{\phi } )]} \nonumber \\
    \equiv&\, \frac{1}{\mathcal{Z}} \int_{ }^{ } D\tilde{\phi} \; 
    \mathcal{O}\left(f^{-1}(\tilde\phi)\right)\, e^{-\tilde{S}[\tilde{\phi}]},
\end{align}
where we have defined the new action \(\tilde{S}\) in the transformed
coordinates to be
\begin{align}
\tilde{S}(\tilde{\phi}) \equiv S(f^{-1}(\tilde{\phi})) - \log\det
J[f^{-1}(\tilde{\phi})].
\label{eq:transformed-action}
\end{align}
If the new probability distribution, \(e^{-\tilde{S}(\tilde{\phi})}\) is
easier to sample than \(e^{-S(\phi)}\) then performing HMC with the new
variables, \(\tilde{\phi }\), would result in a Markov chain \(\{
\tilde{\phi}_{i}\}_{i=1}^{N}\) with lower autocorrelation times for the
observable \(\mathcal{O}\). 

We will refer to this algorithm as Flow HMC (FHMC), and its workflow is as
follows:
\begin{enumerate}
\item Train the network \(f\) by minimising the KL divergence in Equation~(\ref{eq:LossFun}).
\item Run the HMC algorithm to build a Markov chain of configurations using the action $\tilde{S}$,
\begin{align*}
    \{ \tilde{\phi}_{1},\; \tilde{\phi}_{2},\; \tilde{\phi}_{3},\; \dots ,\;
    \tilde{\phi}_{N} \} \sim e^{-\tilde{S}(\tilde{\phi})}.
\end{align*}
\item Apply the inverse transformation \(f^{-1}\) to every configuration in the Markov
chain to undo the variable transformation. This way we obtain a Markov chain
of configurations following the target probability distribution $p(\phi) =
  e^{-S[\phi]}$,
\begin{align*}
    \{ f^{-1}(\tilde{\phi}_{1}),\; f^{-1}(\tilde{\phi}_{2}),\;
    f^{-1}(\tilde{\phi}_{3}),\; \dots ,\; f^{-1}(\tilde{\phi}_{N}) \} \nonumber \\
    =
    \{ \phi_{1},\; \phi_{2},\; \phi_{3},\; \dots ,\; \phi_{N} \} \sim
    e^{-S(\phi)}.
\end{align*}
\end{enumerate}
Note that the HMC acceptance in step 2 can be made arbitrarily high by 
increasing the
number of integration steps in the molecular dynamics evolution of HMC. 
Contrary to what happens in the algorithm suggested in 
Reference~\cite{Albergo2019}, this acceptance does not measure how well $f^{-1}$ approximates a
trivializing map; the relevant question is whether this algorithm improves 
the autocorrelation of HMC.

The motivation behind this work is that a Normalizing Flow parametrised by 
NNs ought to be better able to approximate a trivializing map than the 
leading-order approximation of the flow equation introduced and tested in 
References~\cite{LuscherTrivializingMaps, Schaefer2011}.
Similar ideas have been explored in References~\cite{Foreman2022, Jin2022,
Bacchio2022}. 
Particularly, in Reference~\cite{Jin2022} a Normalizing Flow is optimised 
to approximate the target distribution from an input distribution 
corresponding to the action at a coarser lattice spacing, while training is 
done by minimising the force difference between two theories instead of the 
KL divergence.
In contrast to these previous works we focus on minimal models and cheap training setups in an attempt to avoid the exploding training costs reported in Reference~\cite{DelDebbio:2021qwf}.

% \begin{algorithm}[t]
%     \caption{Flow-based MCMC}
%     \label{alg:flowMCMC}
%     \begin{algorithmic}[1]
%     \Require configuration $\phi$, action $S$, network $f$, probability
%     distribution $p_{f}(\phi)$, trivial distribution $r(z)$
%     \Ensure new configuration $\phi_{\text{new}}$
%     \Function{FlowBasedMCMC}{\phi, S, f}

%     \State{$z$ \sim r(z)}

%     \State{$\phi' = f^{-1}(z)$}

%     \State{\phi_\text{new} \gets \Call{AcceptReject}{\phi', \phi, S, p_f}}

%     \State \Return \phi_\text{new}

%     \EndFunction
%     \end{algorithmic}
% \end{algorithm}

\section{Model and setup}
\label{sec:model-and-setup}

For our study we focus on a $\phi^{4}$ scalar field theory in $D=2$ dimensions
with bare mass $m_{0}$ and bare coupling $g_{0}$. Its standard continuum action
is
\begin{align}
    S[ \phi ] =  \int d^2 x  &\left[ \frac{1}{2} \left( \partial_{\mu} \phi(x)
    \right) \left( \partial_{\mu}\phi(x) \right) \right.\nonumber \\
    &\quad \left. + \frac{1}{2}
m_{0}^{2}\phi(x)^2 + \frac{1}{4!} g_{0} \phi(x)^4 \right].
    \label{eq:phi4-normal-action}
\end{align}
On the lattice we will work with the $\beta$--$\lambda$ parametrization,
\begin{align}
    S[ \phi ] = \sum_{x}^{} \left[ -\beta \sum_{\mu=1}^{2} \phi_{x + e_{\mu}}
    \phi_{x} + \phi_{x}^{2} + \lambda(\phi_{x}^{2} - 1)^2 \right],
    \label{eq:action-beta-lambda}
\end{align}
where $e_{\mu}$ represents a unit vector in the $\mu$-th direction and the sum
runs over all lattice points $x\equiv(x_{1},x_{2})$. The relationship between
these two actions is explained in Appendix~\ref{sec:phi4-param}.

\subsection{Observables}
\label{sec:observables}

We will focus only on a handful of observables, the simplest one being the
magnetization \begin{align}
    M = \frac{1}{V} \sum_{x}^{} \phi_{x},
\end{align}
with $V$ the lattice volume. The building block for the rest of the 
observables is
the connected two-point correlation function
\begin{align}
    G(y) &= \frac{1}{V} \sum_{x }^{} 
    \langle (\phi_{x+y} - \langle \phi \rangle) 
    (\phi_{x} - \langle \phi \rangle) \rangle
    \nonumber \\
         &= \frac{1}{V} \sum_{x }^{} 
         \langle\phi_{x+y}\phi_{x} \rangle - \langle \phi \rangle^2,
\label{eq:2-pt-corr}
\end{align}
where we have used translational invariance to define 
$\langle \phi \rangle =
\langle \phi_x \rangle$. The correlation length,
$\xi$, corresponding to the inverse mass of the lightest mode in the 
spectrum, can be extracted from the spatially-averaged two-point function,
\begin{align}
    \sum_{y_{1}=0}^{L-1} G(y_{1},y_{2}) 
    \propto \cosh \left( \frac{y_{2} - L / 2}{\xi}
    \right)\, ,
\end{align}
at sufficiently large Euclidean time separations $y_{2}$.

We can also measure the one-point susceptibility,\footnote{ 
Note that in other papers \cite{Albergo2019, DelDebbio:2021qwf} the 
susceptibility is defined as $\chi = \sum_{y}^{}G(y)$.}
\begin{align}
    \chi_{0} \equiv G(0) = \frac{1}{V} \sum_{x }^{} \left[ \langle
    \phi_{x}^2 \rangle - \langle \phi \rangle^2 \right].
    \label{eq:1ptsuscep}
\end{align}
Since both the magnetization and the one-point susceptibility are local
observables, we additionally studied the one-point susceptibility measured in
smeared field configurations,
\begin{align}
    \chi_{0,t} \equiv \frac{1}{V} \sum_{x }^{} \left[ \langle
    \phi_{x,t}^2 \rangle - \langle \phi_{t} \rangle^2 \right].
    \label{eq:1ptsuscepflow}
\end{align}
The smeared field configurations $\phi_{t}$ are obtained by solving the 
heat equation
\begin{align}
    \partial_{t}\phi_{x,t} = \partial_{x}^2 \phi_{x,t},
\end{align}
up to flow time $t$, which we choose so that the smearing radius of the flow is
equal to the correlation length of the system, i.e.  $\sqrt{4t} = \xi$. 
For more details, see Appendix~\ref{sec:smearing}.

As usual, the estimator of the expectation values of these observables is the statistical average over the generated Markov chain of
configurations,
\begin{align}
\overline{\mathcal{O}} = \frac{1}{N} \sum_{i=1}^{N} \mathcal{O}(\phi^{(i)}).
\end{align}
where $\mathcal{O}$ is the observable studied. The error associated with this
estimation is given by the statistical variance,
\begin{align}
\sigma_{\overline{\mathcal{O}}}^2 = \frac{\sigma_{\mathcal{O}}^2}{N} 2 \tau_{\text{int},\mathcal{O}},
\end{align}
where \(\tau_{\text{int},\mathcal{O}}\) is the so-called \emph{integrated
autocorrelation time}. It is defined as
\begin{align}
\tau_{\text{int},\mathcal{O}} = \frac{1}{2} + \sum_{m=1}^{\infty}
\frac{\Gamma_{\mathcal{O}}(m)}{\Gamma_{\mathcal{O}}(0)},
\end{align}
with 
\begin{align}
\Gamma_{\mathcal{O}}(m) = {1\over N} \sum_i
\mathcal{O}(\phi^{(i+m)})\mathcal{O}(\phi^{(i)}) - \overline{\mathcal{O}}^2,
\end{align}
being the autocorrelation time of the observable measured at field
configurations separated by \(m\) Markov chain configurations. We estimate
$\tau_{\text{int}}$ using the automatic windowing procedure of the $\Gamma$
method \cite{Wolff:2003sm,Virotta2010}. Particularly, we perform the
autocorrelation analysis with ADerrors.jl~\cite{Ramos2020}, which combines the
$\Gamma$ method with automatic differentiation techniques~\cite{Ramos:2018vgu,Griewank2008}.

\subsection{Network architecture and training}
\label{sec:networkarch}

As mentioned in Section~\ref{sec:intro} we focused on keeping the training 
costs negligible with respect to the cost of producing configurations with 
FHMC. The most intuitive choices that we took for this optimization are:
\begin{enumerate}
    \item
        Use Convolutional Neural Networks (CNNs) instead of fully connected
        networks. The action of Equation~(\ref{eq:action-beta-lambda}) has
        translational symmetry, so the network $f$ should apply the same
        transformation to $\phi_{x}$ for all $x$. CNNs respect this translational
        symmetry, and also require less parameters than fully connected
        networks.
    \item
        Tune the number of layers and kernel sizes of the CNNs so that the
        footprint of the network $f$ is not much bigger than the correlation
        length $\xi$. Two-point correlation functions will generally decay with
        $\sim e^{-\left| x - y \right| / \xi}$, so the transformation of
        $\phi_{x}$ should not depend on $\phi_{y}$ if $\left| x - y \right| \gg
        \xi$. Also, limiting the number of layers reduces the number of
        trainable parameters.
    \item
        Enforce $f$ to satisfy $f(-\phi) = -f(\phi)$. The action in
        Equation~(\ref{eq:action-beta-lambda}) is invariant under $\phi \to -\phi$, so
        enforcing equivariance under this symmetry should optimise training
        costs.
\end{enumerate}

% https://tex.stackexchange.com/questions/353132/align-input-and-output-of-algorithm-to-left
% https://tex.stackexchange.com/questions/163768/write-pseudo-code-in-latex
\begin{figure}
\begin{algorithm}[H]
    \begin{algorithmic}[1]
    % \Require configuration $\phi$, action $S$
    % \Ensure new configuration $\phi_{\text{new}}$
    \Function{hmc}{$\phi$, S}

    \State{$\pi \gets$ \Call{GenerateMomenta}{}}

    \State{$(\phi^{\prime}, \pi^{\prime}) \gets$ \Call{leapfrog}{$\phi$, $\pi$, S}}

    \State{$\phi_\text{new} \gets$ \Call{AcceptReject}{$(\phi', \pi'), (\phi,
    \pi), S$}}

    \State \Return $\phi_\text{new}$

    \EndFunction
    \end{algorithmic}
    \caption{HMC}
        % \\ \footnotesize{The HMC function receives the previous
        %     configuration of the Markov Chain $\phi$ and the action $S$ to be
            % sampled, and returns the next configuration $\phi_{\text{new}}$ of
    % the Markov Chain.}
    \label{alg:hmc}
\end{algorithm}
\end{figure}

\noindent Following \cite{Albergo2019}, we partition the lattice using a
checkerboard pattern, so that each field configuration can be split as
$\phi=\{\phi^A,\phi^B\}$, where $\phi^A$ and $\phi^B$ collectively denote the
field variables belonging to one or the other partition. We then construct the
transformation $f$ as a composition of $n$ layers,
\begin{align}
    f(\phi) = g^{(1)}(g^{(2)}(\dots g^{(n)}(\phi) \dots)),
\end{align}
where each layer $g^{(i)}$ does an affine transformation to a set of the field
variables, $\{\phi^A,\phi^B\}$, organised in a checkerboard
pattern, such as
\begin{align*}
    \phi_{x}^{A} =
    \begin{cases}
        \phi_{x} \quad &\text{if } x_{1}+x_{2} \text{ odd} \\
        0 \quad &\text{otherwise}
    \end{cases},
    \;
    \phi_{x}^{B} =
    \begin{cases}
        0 \quad &\text{if } x_{1}+x_{2} \text{ odd} \\
        \phi_{x} \quad &\text{otherwise}
    \end{cases},
\end{align*}
where $x=(x_{1},x_{2})$. In the affine transformation
\begin{align}
    \label{eq:afftrafo}
    g^{(i)}(\{\phi^A,\phi^B\}) = \{\phi^A, \phi^B \odot e^{\left| s ^{(i)}(\phi^A) \right|} +
    t^{(i)}(\phi^{A})\}
  %  \begin{cases}
    %    z_{a} = \phi_{a} \\
      %  z_{b} = \phi_{b} \odot e^{\left| s _{i}(\phi_{a}) \right|} +
       %t_{i}(\phi_{a})
   % \end{cases},
\end{align}
the partition $\phi^A$ remains unchanged and only the
field variables $\phi^B$ are updated. $s(\phi)$ and $t(\phi)$ are CNNs with
kernel size $k$. To make this transformation equivariant under $\phi \to -\phi$
we enforce $f(-\phi)=-f(\phi)$ by using a tanh activation function and no bias
for the CNNs (see Sec.~III.F of \cite{DelDebbio:2021qwf}). The checkerboard
pattern ensures that the Jacobian matrix has a triangular form so its
determinant can be easily computed, and reads
\begin{align}
    \left| \det \frac{\partial g^{(i)}(\phi)}{\partial \phi} \right| =
    \prod_{\{x | \phi^{B}_{x} = \phi_{x}\}} e^{\left| s ^{(i)}_{x}(\phi^{A}) \right|},
\end{align}
where the product runs over the lattice points where the partition $\phi_{x}^{B} =
\phi_{x}$. An example of the action of a CNN with only 1 layer and a tanh
activation function over a lattice field would be
\begin{align}
    s ^{(i)}_{x}(\phi^{A}) = \tanh \left[ \sum_{y \in \left[
    -\frac{k-1}{2},\frac{k-1}{2} \right]^2 }^{} w^{(i)}(y) \phi_{x-y}^{A}
\right],
\end{align}
where $w^{(i)}(y) \equiv w^{(i)}_{y + (k+1) / 2}$ is the weight matrix of
size $k \times k$ of the CNN $s$ of kernel size $k$. Choosing the same
functional form for $t_{x}^{(i)}(\phi^{A})$, it is easy to check that
the transformation in Equation~(\ref{eq:afftrafo}) is equivariant under
$\{\phi^{A}, \phi^{B}\} \to \{- \phi^{A}, -\phi^{B}\}$.

Two different affine layers with alternate checkerboard patterns are necessary
to transform the whole set of lattice points, and we will denote such a pair of
layers as a \emph{coupling} layer.\footnote{See Reference~\cite{Albergo:2021vyo}
    for an example of an actual implementation of all these concepts.}

In this work we studied network architectures with $N_{l} = 1$ affine coupling
layers, while the CNNs, $s$ and $t$, have kernel size $k$ and no hidden layers.
The output configuration is rescaled with an additional trainable parameter. Finally, 
independent normal distributions are used as the prior distributions $r$ in
Equation~(\ref{eq:EasyProb}).

\subsection{FHMC implementation}

The main focus of this work is the scaling of the autocorrelation times of the
magnetization, $\tau_{M}$, and one-point susceptibilities, $\tau_{\chi_{0}}$ and
$\tau_{\chi_{0,t}}$. Using local update algorithms such as HMC, these
autocorrelation times are expected to scale as
\cite{Baulieu:1999wz}
\begin{align}
  \tau \sim \xi^2.
\end{align}
We will benchmark the scaling of the autocorrelation times in the FHMC algorithm against those in standard HMC.

For a scalar field theory, the HMC equations of motion read
\begin{align}
\dot{\phi}_{x} =\, \pi_{x}, \qquad \dot{\pi}_{x} = -\nabla_{\phi_{x}}S[\phi],
\end{align}
where the force for the momenta $\pi$ follows from the derivative of the action
in Equation~(\ref{eq:action-beta-lambda}),
\begin{align}
    F_{x} \equiv -\nabla_{x} S[ \phi ] &= \beta  \sum_{\mu=\pm1}^{\pm2}
    \phi_{x+e_{\mu}} \\
                                        &\quad+ 2 \phi_{x} \left( 2 \lambda
                                        \left( 1-\phi_{x}^2 \right) -1 \right).
                                        \nonumber
\end{align}
In our simulations we used a leapfrog integration scheme with a single time
scale, and the step size of the integration was tuned to obtain acceptances of
approximately 90\%. A pseudocode of an HMC  implementation is depicted in
Algorithm~\ref{alg:hmc}: the HMC function receives as input a configuration, $\phi$,
and the action of the target theory, $S$; after generating random momenta,
the leapfrog function performs the molecular dynamics step and a
configuration, $\phi_{\text{new}}$, is chosen between the evolved field,
$\phi'$, and the old field, $\phi$, with the usual MH accept--reject step.

\begin{table*}[t!]
    \begin{tabular}{l@{\qquad}
        r@{\qquad}r@{\qquad}r@{\qquad}r@{\qquad}r@{\qquad}r@{\qquad}r@{\qquad}r@{\qquad}r@{\qquad}r}
        \hline
        \hline
        \(L\) & 6 & 8 & 10 & 12 & 14 & 16 & 18 & 20 & 40 & 80\\
        \(\beta\) & 0.537 & 0.576 & 0.601 & 0.616 & 0.626 & 0.634 & 0.641 & 0.645 & 0.667 & 0.677\\
        \(\lambda\) & 0.5 & 0.5 & 0.5 & 0.5 & 0.5& 0.5 & 0.5 & 0.5 & 0.5 & 0.5\\
        \hline
        \hline
    \end{tabular}
    \caption{Studied parameters of the action in
        Equation~(\ref{eq:action-beta-lambda}). $\beta$ has been chosen so that $\xi =
        L / 4$, so the continuum limit is taken in the direction of increasing
    $L$.}
    \label{tb:contlimit}
\end{table*}

\begin{figure}
\begin{algorithm}[H]
    \caption{FHMC}
        % \\ \footnotesize{The FHMC function receives the previous
        %     configuration of the Markov Chain $\phi$, the transformed action
        %     $\tilde{S}$, and the neural network $f$. It returns the next
    %     configuration $\phi_{\text{new}}$ of the Markov Chain.}
    \label{alg:flowhmc}
    \begin{algorithmic}[1]
    % \Require configuration $\phi$, action $S$, network $f$
    % \Ensure new configuration $\phi_{\text{new}}$
    \Function{fhmc}{$\phi, \tilde S, f$}

    \State{$\tilde{\phi} \gets f(\phi)$} \label{algln:phi-phip}

    \State{$\pi \gets$ \Call{GenerateMomenta}{}}

    \State{($\tilde\phi^{\prime}$, $\pi^{\prime}$) $\gets$
    \Call{leapfrog}{$\tilde\phi, \pi, \tilde S$}} \label{algln:flowleapfrog}

    \State{$\tilde\phi_\text{new} \gets$ \Call{AcceptReject}{$(\tilde \phi', \pi'),
            (\tilde \phi,
    \pi), \tilde S$}} \label{algln:flowaccrej}

    \State{$\phi_\text{new} \gets f^{-1}(\tilde \phi_\text{new})$}
    \label{algln:phip-phi}

    \State \Return $\phi_\text{new}$

    \EndFunction
    \end{algorithmic}
\end{algorithm}
\end{figure}

The proposed FHMC algorithm is in essence the HMC algorithm with the transformed
action in Equation~(\ref{eq:transformed-action}), which arises from the change of
variables $\tilde{\phi} = f(\phi)$ in Equation~(\ref{eq:pathintegral-chvar}). The new
Hamilton equations of motion now include derivatives with respect to the new
variables $\tilde{\phi}$,
\begin{align}
\label{eq:flow-Heqs}
\dot{\tilde{\phi}}_{x} =\, \pi_{x}, \qquad \dot{\pi}_{x} =
-\nabla_{\tilde{\phi}_{x}}\tilde{S}[\tilde{\phi}].
\end{align}
The basic implementation is sketched in Algorithm~\ref{alg:flowhmc}. The main
differences with respect to standard HMC in  Algorithm~\ref{alg:hmc} are line
\ref{algln:phi-phip}, where we transform from the variables $\phi$ to the
variables $\tilde{\phi}$ using the trained network $f$; and line
\ref{algln:phip-phi}, where we undo the change of variables to obtain the new
configuration $\phi_{\text{new}}$ from $\tilde{\phi}_{\text{new}}$. 

Note that the molecular dynamics evolution and the accept--reject step,
lines \ref{algln:flowleapfrog} and \ref{algln:flowaccrej}, are applied to the
transformed field variables $\tilde{\phi}$ with the new action $\tilde{S}$. 
Irrespective of the transformation $f$, the acceptance rate can be made arbitrarily high by reducing numerical errors in the integration of the equations of motion in Equation~(\ref{eq:flow-Heqs}).
This means that we will always be able to tune the FHMC acceptance to approximately 90\% by increasing the number of integration steps, even for a poorly trained Normalizing Flow.

Note also that now the evaluation of the force $\tilde{F}_{x} \equiv -
\nabla_{\phi_{x}}\tilde{S}[ \tilde{\phi} ]$ requires computing the derivative of
$\log \det J[ f^{-1}(\tilde{\phi}) ]$. This cannot be written analytically for
an arbitrary network, and we used PyTorch's automatic differentiation methods
for its evaluation \cite{NEURIPS2019_bdbca288}.

% \begin{table}[t!]
%     \caption{Studied parameters of the action in Eq.
%         \ref{eq:action-beta-lambda}. $\beta$ has been chosen so that $\xi = L / 4$,
%     so the continuum limit is taken in the direction of increasing $L$.}
%     \begin{array}{lrrrrrrrrrr}
%         \hline
%         \hline
%         \(L\) & 6 & 8 & 10 & 12 & 14 & 16 & 18 & 20 & 40 & 80\\
%         \(\beta\) & 0.537 & 0.576 & 0.601 & 0.616 & 0.626 & 0.634 & 0.641 & 0.645 & 0.667 & 0.677\\
%         \(\lambda\) & 0.5 & 0.5 & 0.5 & 0.5 & 0.5 & 0.5 & 0.5 & 0.5 & 0.5 & 0.5\\
%         \hline
%         \hline
%     \end{array}
%     \label{tb:contlimit}
% \end{table*}

% \begin{table}[t]
%     \begin{tabular}{lrrrrrrrrrr}
%         \hline
%         \hline
%         \(L\) & 6 & 8 & 10 & 12 & 14 \\
%         \(\beta\) & 0.537 & 0.576 & 0.601 & 0.616 & 0.626 \\
%         \(\lambda\) & 0.5 & 0.5 & 0.5 & 0.5 & 0.5\\
%         \hline
%         \(L\) & 16 & 18 & 20 & 40 & 80\\
%         \(\beta\) & 0.634 & 0.641 & 0.645 & 0.667 & 0.677\\
%         \(\lambda\) & 0.5 & 0.5 & 0.5 & 0.5 & 0.5\\
%         \hline
%         \hline
%     \end{tabular}
%     \caption{Studied parameters of the action in
%         Equation~(\ref{eq:action-beta-lambda}). $\beta$ has been chosen so that $\xi =
%         L / 4$, so the continuum limit is taken in the direction of increasing
%     $L$.}
%     \label{tb:contlimit}
% \end{table}

\section{Results}
\label{sec:orgbc47492}

We are interested in the scaling of the cost towards the continuum limit.
Following the analysis in \cite{DelDebbio:2021qwf}, we tuned the coupling
$\beta$ so that the correlation length satisfies
\begin{align}
    \xi \approx \frac{L}{4}.
\end{align}
The continuum limit is therefore approached in the direction of increasing $L$.
In Table~\ref{tb:contlimit}  we summarise the parameters used in our simulations
of FHMC and standard HMC. Results obtained with both HMC and FHMC can be found
in tables~\ref{tab:hmcresults} and \ref{tb:flowhmc-k3} of
Appendix~\ref{sec:supplementary-tables-plots}, respectively.

\subsection{Minimal network}
\label{sec:minimal-arch}

\begin{figure*}[ht!]
    \centering
    \includegraphics[width=0.49\linewidth]{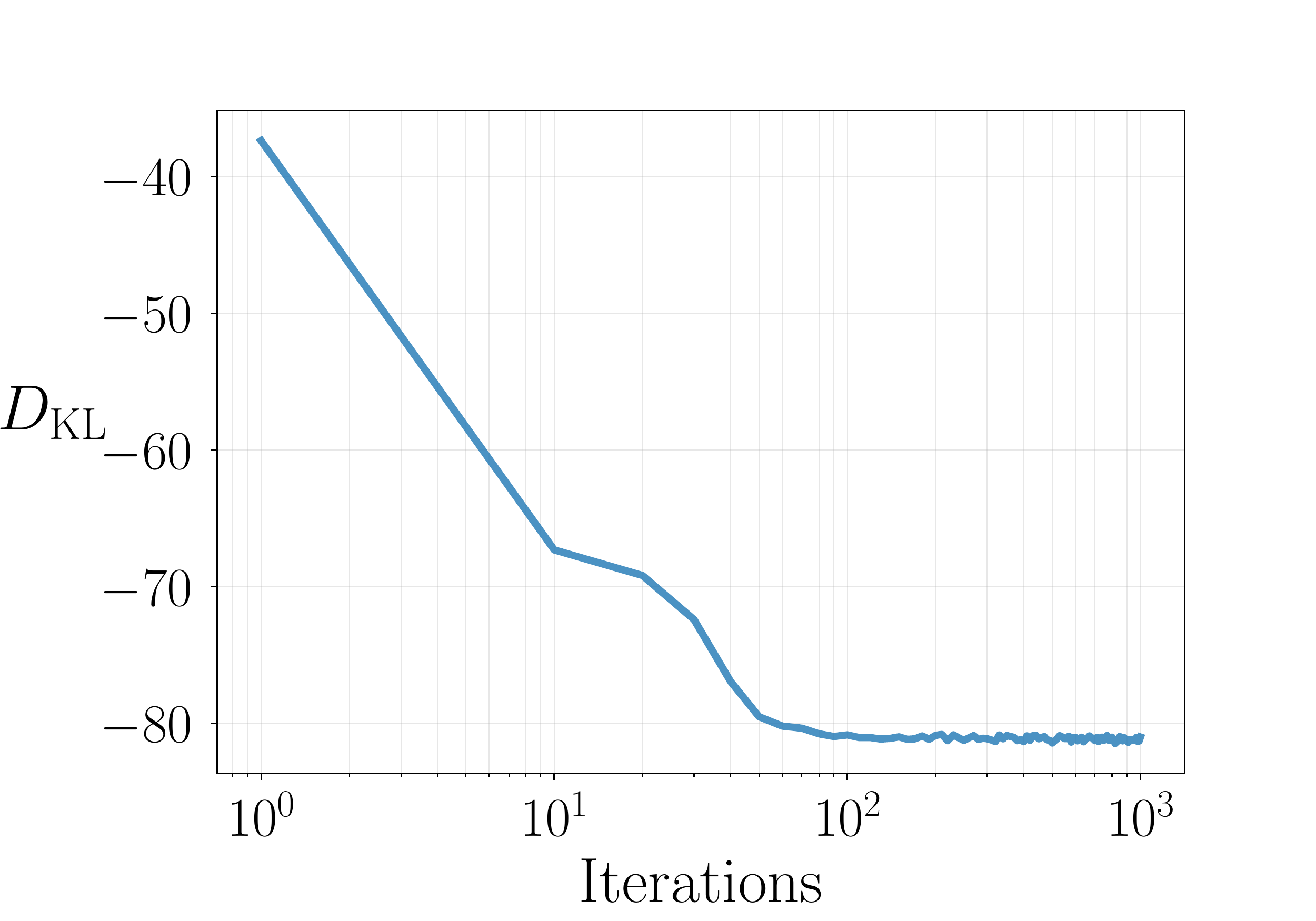}
    \includegraphics[width=0.49\linewidth]{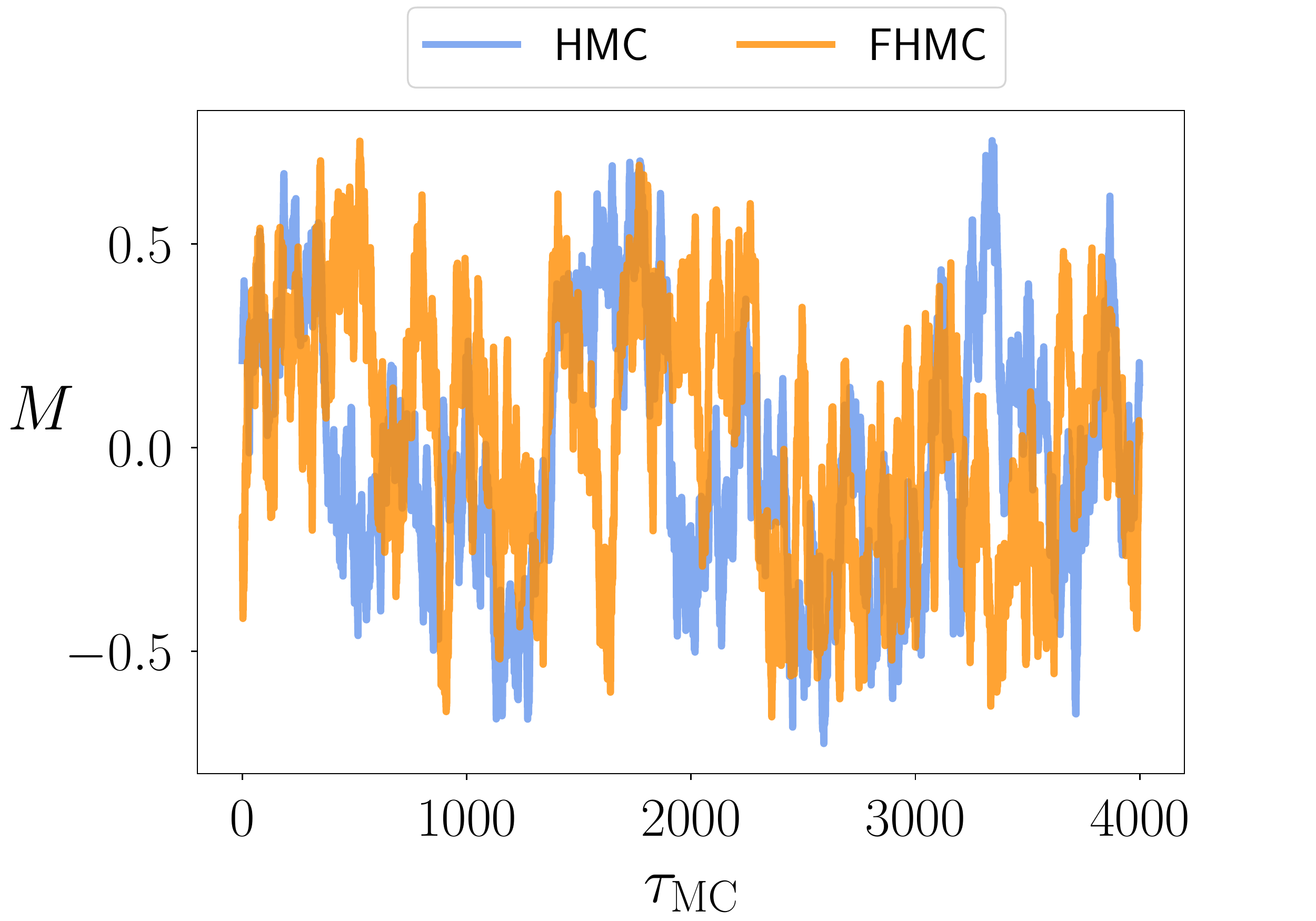}
    \caption{(Left) History of the KL divergence during the training from
        independent Gaussians to a theory with parameters $\beta = 0.641$, $\lambda
    = 0.5$, lattice size $L = 18$ and $\xi = L / 4$. (Right) History of the
magnetization for a simulation with HMC (blue) and FHMC (orange).}
    \label{fig:KLandMhist}
\end{figure*}

\begin{figure}
    \centering
    \includegraphics[width=\linewidth]{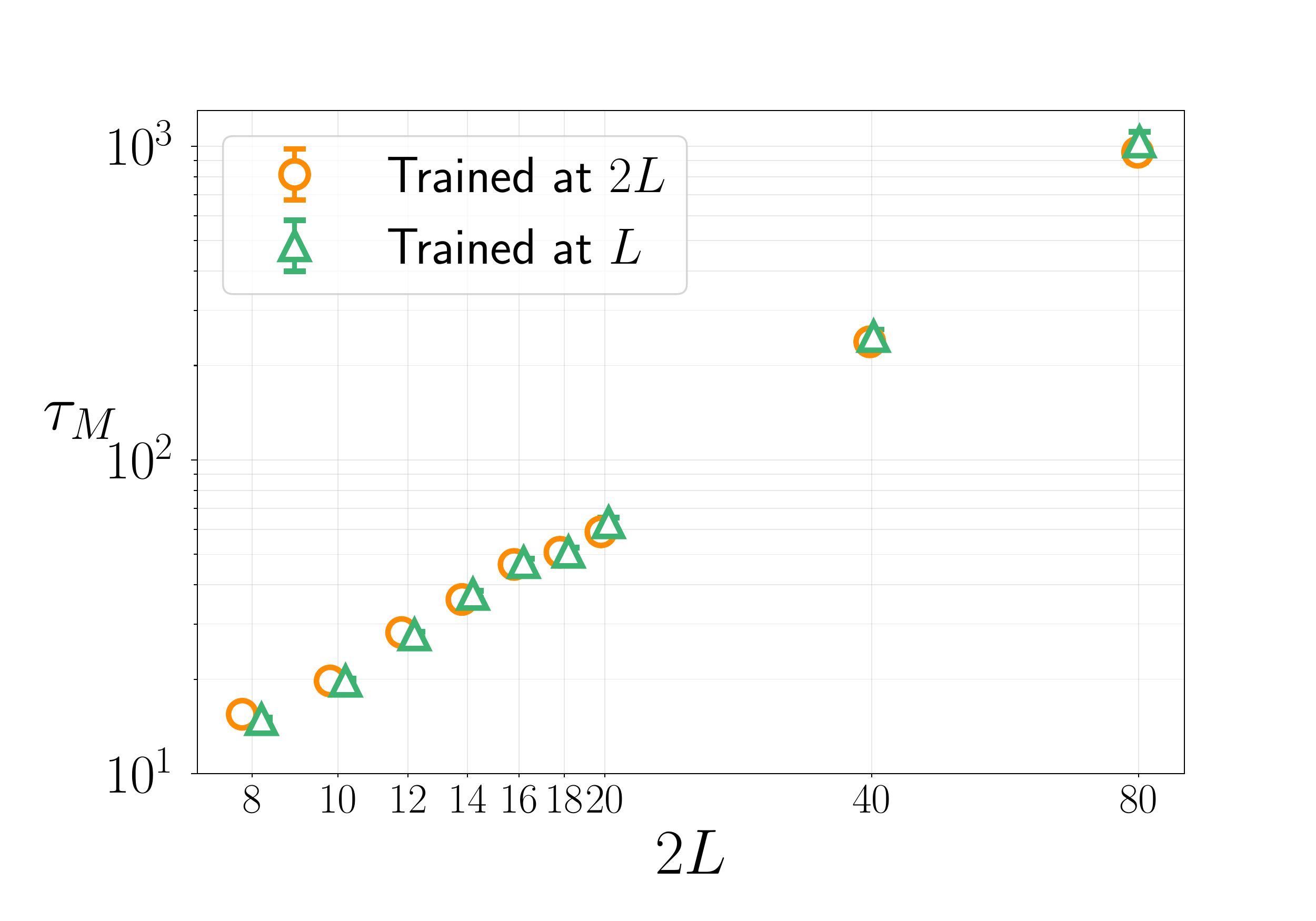}
    \caption{Autocorrelation time of the magnetization at lattice size
        $2L$ using FHMC. In circles, the networks used were
        trained at lattice size $2L$; in triangles, they were trained at
    $L$ and used at $2L$.}
    \label{fig:biggerV}
\end{figure}

\begin{figure*}[!t]
    \centering
    \includegraphics[width=0.49\linewidth]{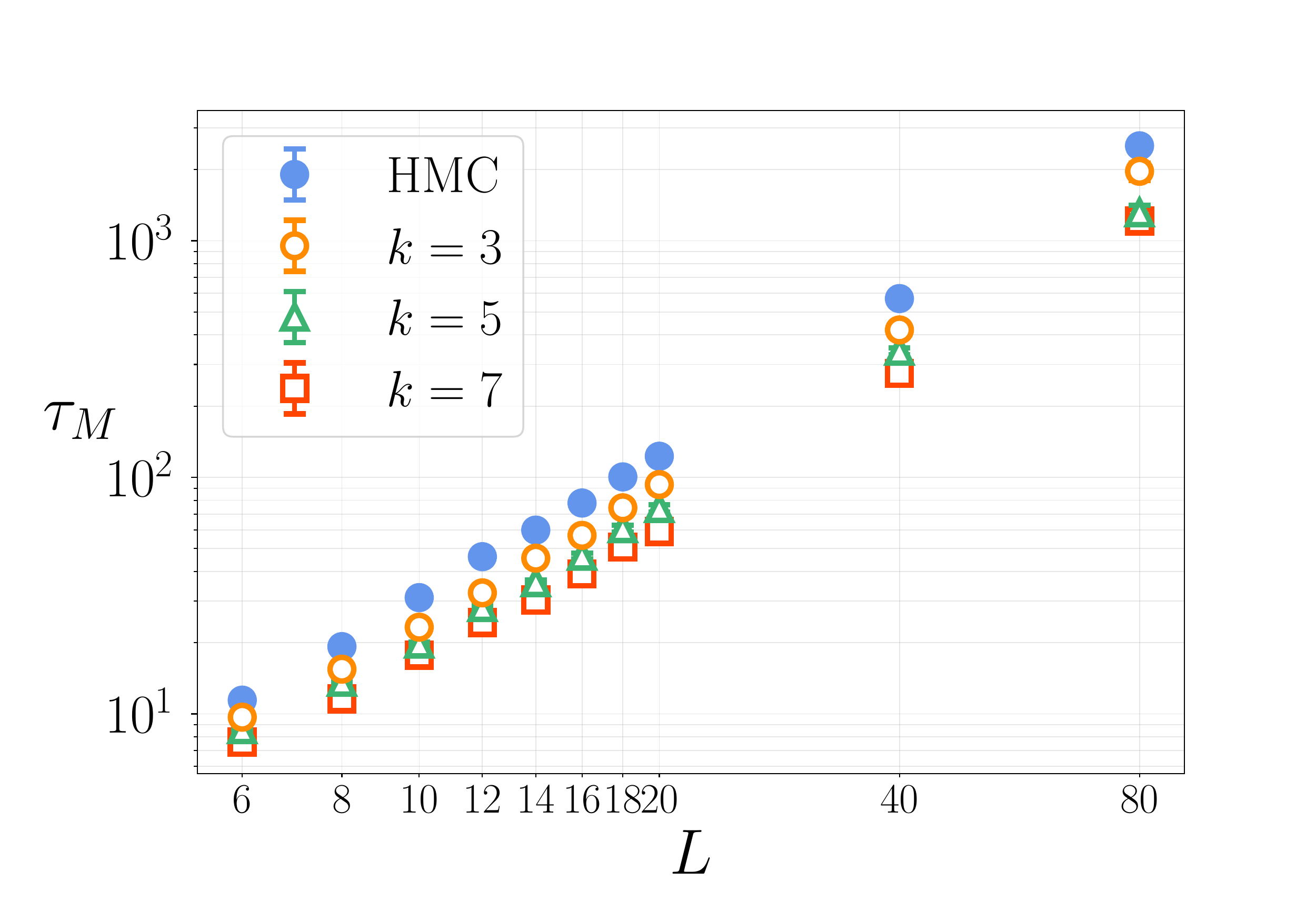}
    \includegraphics[width=0.49\linewidth]{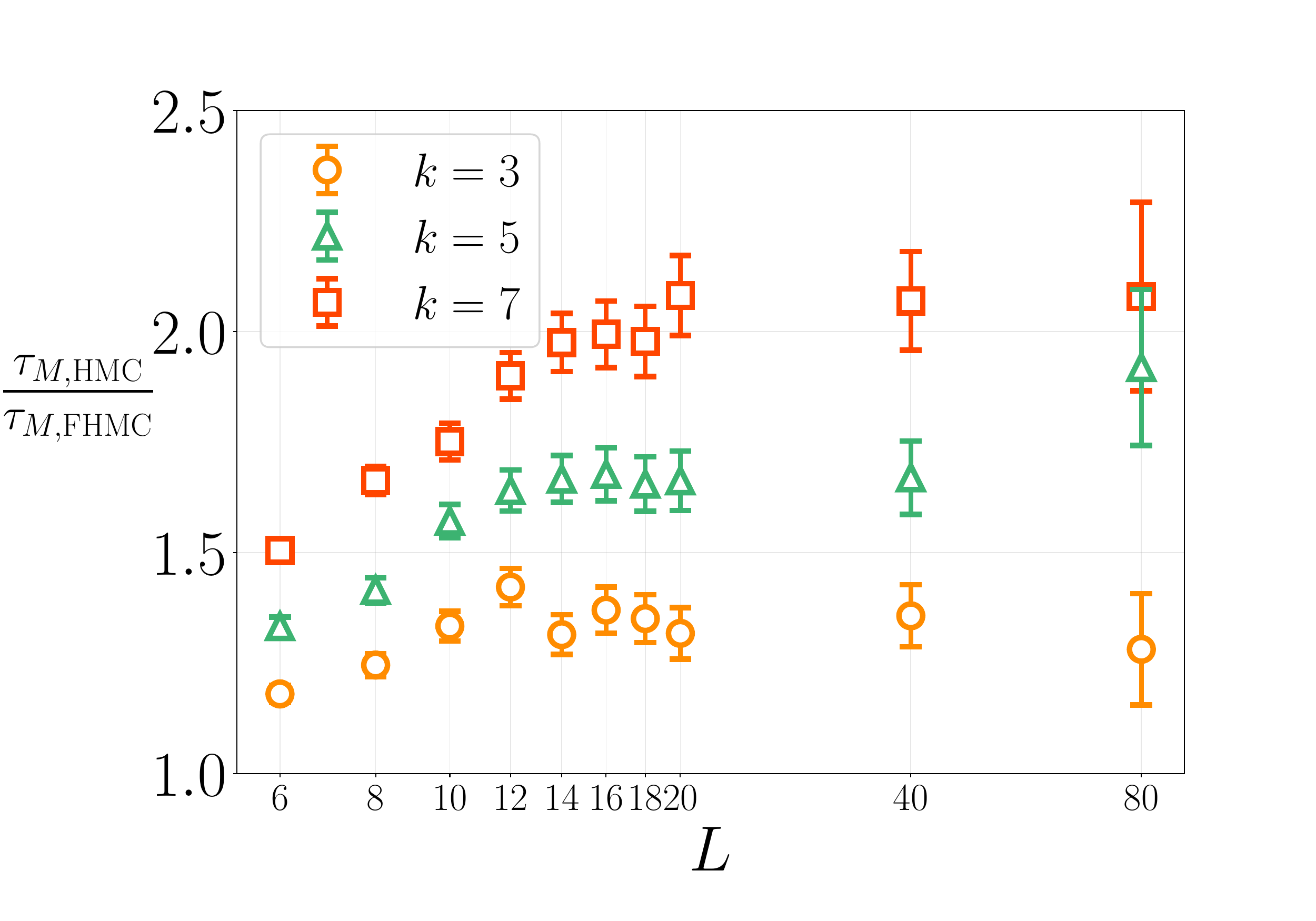}
    \caption{(Left) Scaling of the autocorrelation time of the magnetization
        towards the continuum for HMC (filled, blue circles) and FHMC with
        kernel sizes $k=3,5,7$ (open circles, triangles and squares).  (Right)
        Scaling of the ratio of autocorrelation times of the magnetization of
    HMC with respect to FHMC.} 
    \label{fig:Mfixedarch}
\end{figure*}

An obvious strategy to reduce training costs is to build networks with few
parameters to train.  As mentioned in Section~\ref{sec:model-and-setup}, using
$N_{l} = 1$ coupling layer would suffice to transform the whole lattice, while a
kernel size $k=3$ for the CNNs, which would couple only nearest neighbour, is
the smallest that can be used.\footnote{ The transformation with $k=1$ being a trivial rescaling.} 
Such a network has only 37 trainable parameters in total,\footnote{A CNN
    with 1 layer has $k^2$ parameters.  The transformation in
    Equation~(\ref{eq:afftrafo}) with $N_{l}$ affine coupling layers has
    $4\times N_{l}$ different CNNs, and therefore $4 \times N_{l} \times k^2$
    parameters.  Since we also add a global rescaling parameter as a final layer
    of our network, this architecture has $N_{p} = 4 \times N_{l} \times k^2 +
    1$ parameters.}
and is the most minimal network that we will consider.

It is interesting to study whether such a simple network can learn physics of
a target theory with a non-trivial correlation length. In
Figure~\ref{fig:KLandMhist} (left) we plot the evolution of the KL divergence
during the training of a network with such minimal architecture, where the
target theory has parameters $\beta = 0.641$, $\lambda = 0.5$, lattice size
$L=18$ and correlation length $\xi = L / 4 = 4.5$. Since the network has very
few parameters, the KL divergence reaches saturation after only
$\mathcal{O}(100)$ iterations.

Once the network is trained, one can use it as a variable transformation for the
FHMC algorithm as sketched in Algorithm~\ref{alg:flowhmc}. In
Figure~\ref{fig:KLandMhist} (right) we show the magnetizations of a slice of 4000
configurations of Markov chains coming from FHMC and HMC simulations, yielding
as autocorrelation times
\begin{align}
    \tau_{M, \text{FHMC}} = 74.4(3), \quad
    \tau_{M, \text{HMC}} = 100.4(2).
\end{align}
All the results with this minimal network can be found in
Table~\ref{tb:flowhmc-k3} of Appendix~\ref{sec:supplementary-tables-plots},
showing that FHMC leads to smaller autocorrelations compared to standard HMC,
especially as the continuum limit is approached. This fact seems to indicate
that a network with few parameters can indeed learn transformations with
relevant physical information. 

A measure of the closeness of the distributions $p_f$ and $p$ is given by the MH
acceptance\footnote{Not to be confused with the HMC and FHMC acceptances, which
are tuned to 90\% in this work.}
when sampling $p$ with configurations drawn directly from $p_f$.
Focusing on the first three columns of Table~\ref{tab:infinite-results}, it is
clear that the acceptances are low and decrease towards the continuum limit, in
spite of the fact that the autocorrelation time of FHMC is better than that of
HMC. This is because the networks used have a very reduced set of parameters
and therefore limited expressivity. The map defined via the trained networks is
not very accurate in generating the probability distribution of the target
theory. However FHMC, i.e. a molecular dynamics evolution using flowed
variables, yields a clear gain in the autocorrelation times.

\subsection{Infinite volume limit}

\begin{table}[t]
    \centering
    \begin{tabular}{rrll}
        $L$ & $\beta$ & $\text{Acc. at } L$ & $\text{Acc. at } 2L$\\
        \hline
        3 & 0.537 & 0.3 & 0.2\\
        4 & 0.576 & 0.04 & 0.001\\
        5 & 0.601 & 0.002 & 0.00003\\
        6 & 0.616 & 0.002 & 0.000007\\
        7 & 0.626 & 0.0001 & $< 10^{-7}$\\
        8 & 0.634 & 0.0001 & -\\
        9 & 0.641 & 0.00007 & -\\
        10 & 0.645 & 0.00004 & -\\
    \end{tabular}
    \caption{MH acceptances of networks trained at lattice size $L$ and used to
    sample a theory with coupling $\beta$ at lattice sizes $L$ and $2L$.}
    \label{tab:infinite-results}
\end{table}

An important advantage of using a translationally-invariant network architecture, such as the one
containing CNNs, is that they can be trained at a small lattice size $L$ and then used in a larger 
lattice $L' > L$. Note that doing this in the approach of
References~\cite{Albergo2019, DelDebbio:2021qwf} would not be viable since it would
lead to an exponential decrease in the MH acceptance due to the extensive
character of the action. 

In the last column of Table~\ref{tab:infinite-results} we show the MH
acceptance using networks with the minimal architecture of
Section~\ref{sec:minimal-arch} trained at lattice size $L$ when the target
theory has lattice size $2 L$. One can see that the acceptances are
significantly lower than those obtained with the target theory at size
$L$.  However, the acceptance of the FHMC algorithm
(Algorithm~\ref{alg:flowhmc}) can be kept arbitrarily high by increasing the
number of integration steps of the Hamilton equations, so reusing the networks
for higher volumes does not pose any problem. The reduced MH acceptance does not
translate into a change in the autocorrelation time.  

In Figure~\ref{fig:biggerV} we compare the autocorrelation times for the
magnetization of networks trained at $L$ and reused at $2L$ with the ones of
networks trained directly at $2L$. Since they agree within statistical
significance, this indicates that the relevant physical information is already
learned at small volumes and reinforces the intuition that the training does not need to 
be done at a lattice sizes larger than $\xi$.

\subsection{Continuum limit scaling with fixed architecture}

Finally we want to determine whether the computational cost of FHMC has a better
scaling than the standard HMC as we approach the continuum. First we will consider a
fixed network architecture as we scale $L$. We have trained a different network
for each of the lattice sizes in Table~\ref{tb:contlimit}, with  $N_{l}=1$ and
kernel sizes $k=3,5,7$. The cost of the training is in all cases negligible
with respect to the costs of the FHMC, and the integration step of the leapfrog
scheme is tuned to have an acceptance rate of approximately 90\% for every
simulation, as we do for HMC. 

In Figure~\ref{fig:Mfixedarch} (left) we show the autocorrelation times of the
magnetization for both HMC (filled, blue circles) and FHMC with kernel sizes
$k=3,5,7$ (open circles, triangles and squares). One can see that the
autocorrelations in FHMC are lower than the ones of HMC, and decrease as 
the kernel size of the CNNs is increased.\footnote{ Note that an accurate cost
comparison should include the overhead of computing the force via automatic
differenciation. We have not tried to optimize this step and therefore postpone
a detailed cost comparison to future work.} 

In order to study the scaling towards the continuum limit, we plot 
the ratio of autocorrelation times for HMC versus 
FHMC in Figure~\ref{fig:Mfixedarch} (right) for the three values of the kernel
size. Although for the coarser lattices the ratio increases towards the
continuum, it seems to saturate within the range of lattice spacings explored,
indicating that the cost scaling of both algorithms is the same. The same
behaviour is observed for the one-point susceptibilities of
Equations~(\ref{eq:1ptsuscep}) and (\ref{eq:1ptsuscepflow}).

\subsection{Continuum limit scaling with $k \sim \xi$}
\label{sec:scale-k}

\begin{figure*}[!t]
	\centering
    \includegraphics[width=0.49\linewidth]{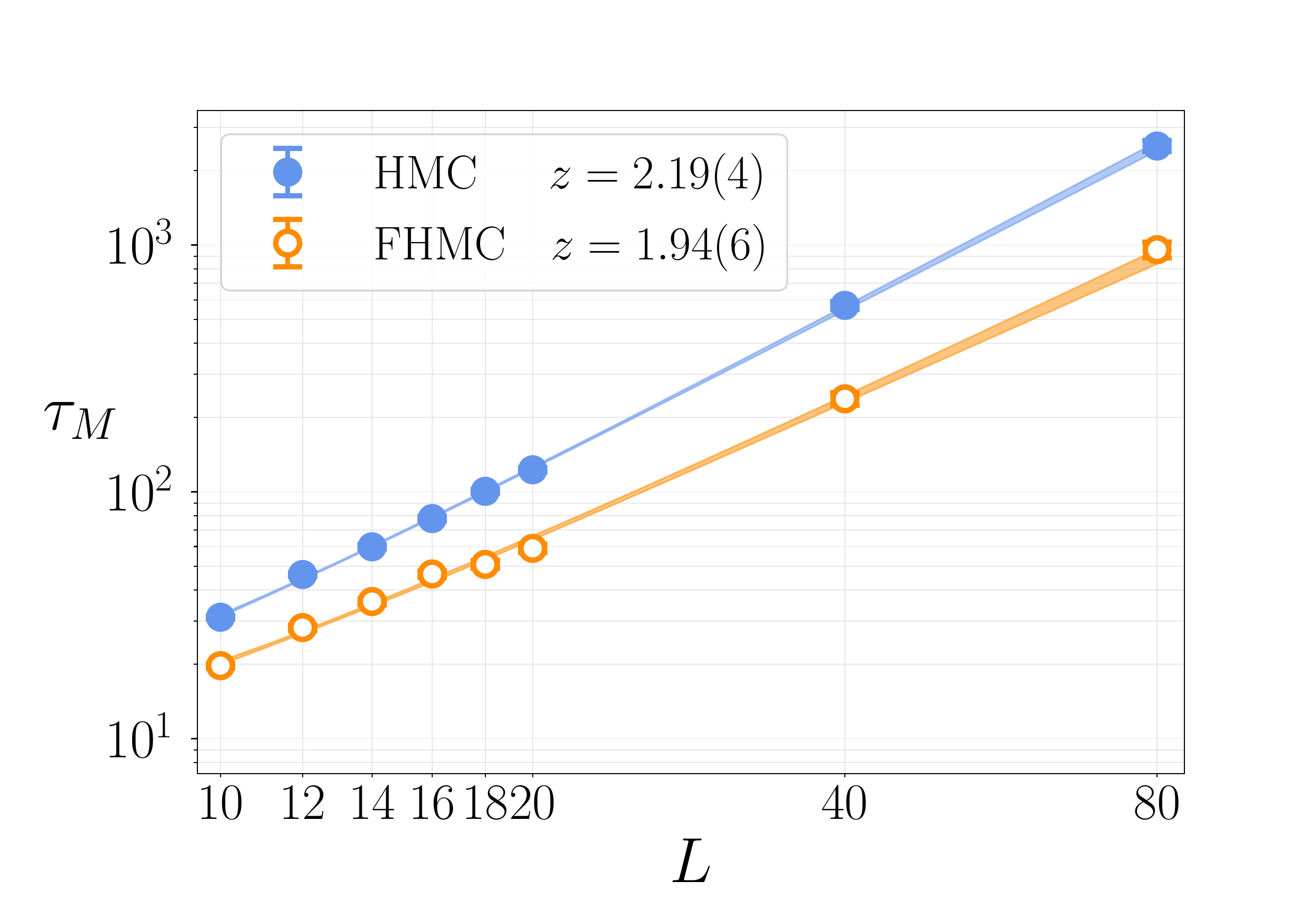}
    \includegraphics[width=0.49\linewidth]{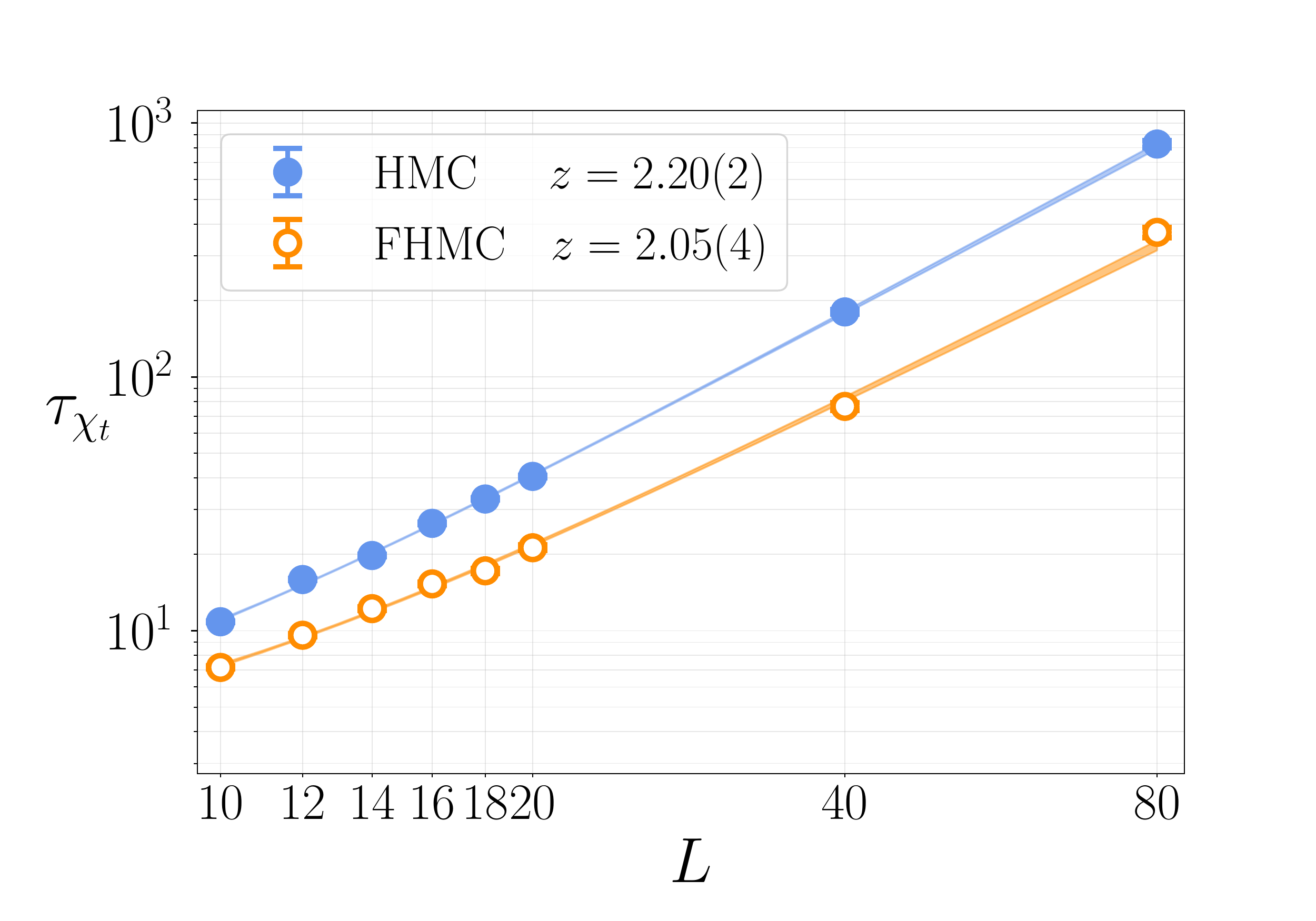}
    \caption{(Left) Scaling of the autocorrelation time of the magnetization for HMC and FHMC with a network with 
        $k\sim\xi$. (Right) Same for the flowed one-point susceptibility,
        Equation~(\ref{eq:1ptsuscepflow}). The fits correspond to
    the best fit function of Equation~(\ref{eq:xiscaling}).}
    \label{fig:tauM-scalingk}
\end{figure*}

As we take the continuum limit the correlation length increases in lattice
units. If the footprint is chosen to scale with $\xi$, the convolution
implemented by the network covers the same physical region. Using our
architecture, this can be done by adding more coupling layers or increasing the
kernel size, $k$. More concretely, a kernel size $k$ couples $(k-1) / 2$ nearest
neighbours; therefore, since we have no hidden layers, if we have $N_{l}$
coupling layers the network will couple $N_{l}(k-1)$ nearest neighbours.

In Figure~\ref{fig:tauM-scalingk} (left) we show again the scaling of the
autocorrelation times of the magnetization, but now the networks used for the
FHMC algorithm have $N_{l}(k-1) \approx \xi$. Particularly, all networks of the
plot have $N_{l} = 1$ coupling layers, so only the kernel size varies: for $L =
10$ to $L = 16$ the networks have $k = 5$; for $L = 18$ and $L=20$, $k=7$; for
$L = 40$, $k = 11$; and for $L = 80$, $k = 21$.

The curves of the plot correspond to fits to the function
\begin{align}
    \tau = a \xi^{z} + b,
       \label{eq:xiscaling}
\end{align}
yielding as result
\begin{align}
    z_{M, \text{HMC}} =\; 2.19(4), \quad z_{M, \text{FHMC}} =\; 1.94(6).
    \label{eq:flowscaling-k}
\end{align}
Thus keeping the physical footprint size constant seems to yield a slight
improvement in the scaling towards the continuum.\footnote{As a word of caution,
    the errors in the results of Equation~(\ref{eq:flowscaling-k}) shall not be
interpreted as Gaussian; the error of $\tau_{\text{int}}$ involves a sum over
the four-point autocorrelation function, whose computation is
usually approximated~\cite{Madras1988,Luscher2004,Virotta2010}.} It is also
interesting to
see that the same happens with the smeared susceptibility in
Figure~\ref{fig:tauM-scalingk} (right). The latter  is a
non-local observable that has been measured in smeared configurations with
a smoothing radius $\sqrt{4t} = \xi$ (see Appendix~\ref{sec:smearing}).

This slight improvement is in agreement with the fact that for a fixed
network architecture the continuum scaling remains the same as HMC, while
increasing the kernel size improves the global factor of the autocorrelations,
as was seen in Figure~\ref{fig:Mfixedarch}.

\begin{figure}[!t]
	\centering
    \includegraphics[width=\linewidth]{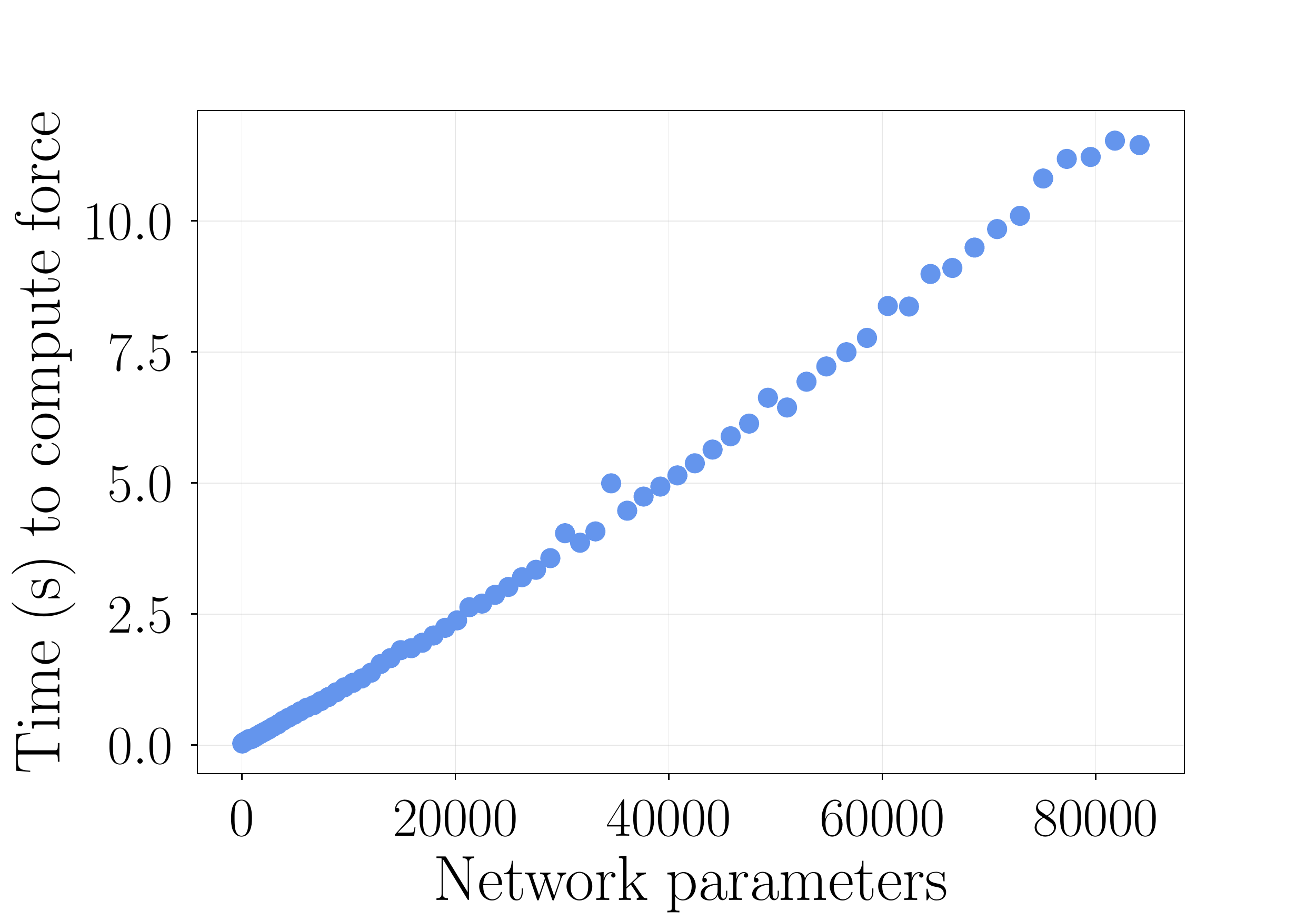}
    \caption{Dependence of the time required  to evaluate the force on the
    number of parameters in the network for a lattice with $L=320$.}
    \label{fig:ADscaling}
\end{figure}

It is important to note that increasing the kernel size of the network implies increasing
 the number of parameters in the training, and also the number of operations to compute the force
in the molecular dynamics evolution using automatic differentiation.
Particularly, the number of parameters of our networks is given by
\begin{align}
N_{\text{params}} = 4 k^2 N_{l} + 1.
\end{align}
In Figure~\ref{fig:ADscaling} we show the time needed to compute the force
on a lattice with fixed length $L=320$ as a function of the number of network
parameters (keeping $N_{l} = 1$ and varying $k$ in the interval $k \in
[3,161]$).  Since the computing time seems to scale linearly with the number of
parameters, if $k$ scales with the correlation length $\xi$ then there is an
additional term proportional to $\xi^2$ in the FHMC cost. 

According to this estimate, FHMC would not reduce the asymptotic simulation cost
of a $\phi^4$ theory with respect to HMC. However, the implementation of the
FHMC force does not require the integration of the flow equation in
Equation~(\ref{eq:triv-flow-eq}), unlike in \cite{Schaefer2011}. Knowing that
FHMC already reduces autocorrelation times with minimal architectures, it could
probably be used to reduce simulation costs in Lattice QCD with minimal
implementation effort. 

\section{Conclusions}

We have tested a new algorithm, Flow HMC (FHMC), that implements the
trivializing flow algorithm of References~\cite{Luscher:2010iy,Schaefer2011} via a
convolutional neural network, similar to those used in the Normalizing Flow
algorithms introduced in Reference~\cite{Albergo2019}. In contrast with previous
works on Normalizing Flows, we use minimal network architectures which leads to
negligible training costs, not affected therefore by the bad scaling towards the
continuum limit observed in \cite{DelDebbio:2021qwf}. The main new ingredient is
the combination of a neural network implementation of the trivializing flow with
an HMC integration which keeps a large acceptance for any network architecture. 

We have tested the algorithm in a scalar theory in 2D and benchmarked it against
standard HMC. We have observed a significant reduction of the autocorrelation
times in FHMC for the observables measured: the magnetization and one-point
susceptibility. This improvement is maintained as the physical volume is
increased at fixed lattice spacing, meaning that the network can be trained a
small physical volume $L$ and used at larger one $L'> L$ without any extra cost.
For gauge theories, this opens up the possibility of doing the training at
unphysical values of the quark masses or small volumes --- or both, where
training is cheaper --- and reusing the trained networks to sample the targeted
theory at the physical values of the parameters.

However, for a fixed network architecture the scaling of the autocorrelation
with the lattice spacing remains  the same as that of HMC. A slight improvement
in the scaling of the autocorrelation time is observed when the footprint of the
network is kept constant in physical units. Although the training cost still
remains negligible, scaling the footprint with the correlation length  implies
an extra cost in the computation of the force in the FHMC, leading to a worse
overall scaling than HMC in the theory considered. This might be different in a
theory with fermions, where the dominant cost is the inversion of the Dirac
operator. Also, as discussed in Reference~\cite{Abbott2022b}, the use of other
Machine Learning training techniques such as transfer learning and the use of
optimal architectures and stopping criteria can help alleviating the training
cost scaling.

Although the improvement observed in the autocorrelation times
for fixed network architectures of FHMC might be of some practical use, particularly given
the simplicity of the implementation, it remains to be demonstrated that a
neural network training policy can be applied that avoids critical slowing down.

\section*{Acknowledgments}
\addcontentsline{toc}{section}{Acknowledgements}

We acknowledge support from the Generalitat Valenciana grant PROMETEO/2019/083,
the European projects H2020-MSCA-ITN-2019//860881-HIDDeN and
101086085-ASYMMETRY, and the national project PID2020-113644GB-I00. AR
acknowledges financial support from Generalitat Valenciana through the plan GenT
program (CIDEGENT/2019/040).  DA acknowledges support from the Generalitat
Valenciana grant ACIF/2020/011. JMR is supported by STFC grant ST/T506060/1. LDD
is supported by the UK Science and Technology Facility Council (STFC) grant
ST/P000630/1.

This work has been performed under the Project HPC-EUROPA3
(INFRAIA-2016-1-730897), with the support of the EC Research Innovation Action
under the H2020 Programme; in particular, we gratefully acknowledge the support
of the computer resources and technical support provided by EPCC. This work used
the ARCHER2 UK National Supercomputing Service
(\url{https://www.archer2.ac.uk}).  We also acknowledge the computational
resources provided by Finis Terrae II (CESGA), Lluis Vives (UV), Tirant III
(UV). The authors also gratefully acknowledge the computer resources at
Artemisa, funded by the European Union ERDF and Comunitat Valenciana, as well as
the technical support provided by the Instituto de Física Corpuscular, IFIC
(CSIC-UV).

\bigskip

\appendix

\section{$\phi^4$ theory on the lattice}
\label{sec:phi4-param}

Discretising the Laplacian as
\begin{align}
    \partial^2 \phi_{x} \to \frac{1}{a^2} \sum_{\mu=1}^{2} \left(
    \phi_{x+ae_{\mu}} + \phi_{x-ae_{\mu}} - 2\phi_{x} \right)
\end{align}
and using the translational invariance of the action in
Equation~(\ref{eq:phi4-normal-action}) leads to
\begin{align}
    S(\phi) = \sum_{x}^{} \left[ - \sum_{\mu=1}^{2}\phi_{x}\phi_{x+e_{\mu}} +
    \left( 2+ \frac{m_{0}^{2}}{2} \right) \phi_{x}^2 +
\frac{g_0}{4!}\phi_{x}^{4} \right]. 
\end{align}
Equation~(\ref{eq:action-beta-lambda}) can be obtained with the
transformations
\begin{align}
    \phi \to \sqrt{\beta}\phi, \quad 2 + \frac{m_{0}^{2}}{2} =
    \frac{1-2\lambda}{\beta}, \quad \frac{g_0}{4!} = \frac{\lambda}{\beta^2}.
\end{align}

\section{Smearing}
\label{sec:smearing}

\subsection{Discrete heat equation in $D=2$}

The time evolution of the gradient flow is given by the heat equation
\begin{align}
\partial_{t} \phi_{x,t} = \partial_{x}^2 \phi_{x,t}\;,
\label{eq:he}
\end{align}
where the lattice discretization of the Laplacian is
\begin{align}
\partial_{x}^2 \phi_{x,t} = \frac{1}{a^2} \sum_{\mu=1}^{2} \left[
\phi_{x+e_{\mu},t} + \phi_{x-e_{\mu},t}  - 2\phi_{x,t}\right].
\end{align}
We can solve the gradient flow exactly using the Discrete Fourier
Transform\footnote{ Note that $x \equiv(x_{1}, x_{2})$ and $p \equiv(p_{1},
p_{2})$. For summations we use the shorthand notation $\sum_{x} \equiv
\sum_{x_{1},x_{2}=0}^{(L-1, L-1)}$.} (DFT) and the inverse DFT (IDFT):
\begin{align}
\phi_{p,t} &= \sum_{x} e^{-ipx} \phi_{x,t} \quad && \text{(DFT)}\\
\phi_{x,t} &= \frac{1}{L^2} \sum_{p} e^{ipx} \phi_{p,t} \quad && \text{(IDFT)}
\end{align}
Using this, Equation~(\ref{eq:he}) becomes
\begin{align}
&\frac{1}{L^2} \sum_{p}^{} e^{ipx} \partial_{t}\phi_{p,t} = \partial_{x}^2
\frac{1}{L^2} \sum_{p}^{} e^{ipx} \phi_{p,t} \\ 
&= \frac{1}{L^2}
\sum_{p}^{} \sum_{\mu=1,2}^{}[e^{ip(x+e_{\mu})} + e^{ip(x-e_{\mu})} - 2e^{ipx}]
\phi_{p,t} \; .\nonumber
\end{align}
Then, the expression in square brackets is
\[
    e^{ipx} \times 2[ \cos(p_{\mu})-1]
\]
and therefore
\begin{align}
\partial_{t}\phi_{p,t} = - \sum_{\mu}^{}4 \sin^2 \left( \frac{p_{\mu}}{2} \right)
\phi(p,t)  \equiv - \hat{p}^2 \phi_{p,t} \; ,
\end{align}
where we have defined $p_{\mu} \equiv p \hat{e}_{\mu}$, with $p_{\mu} = 2\pi n /
L$ for $n = 0, \dots, L-1$; and $\hat{p}^2 \equiv \sum_{\mu}^{} 4 \sin^2 \left( \frac{p_{\mu}}{2} \right)$.
The solution of this equation in momentum space is
\begin{align}
\phi_{p,t} = \phi_{p,0} e^{-\hat{p}^2 t} \; ,
\label{eq:he-mom-sol}
\end{align}
which we can express in position space using the IDFT,
\begin{align}
\phi_{x,t} &= \frac{1}{L^2} \sum_{p}^{} e^{ipx} \phi_{p,t} = \frac{1}{L^2}
\sum_{p}^{} e^{ipx}e^{-\hat{p}^2 t} \phi_{p,0} \nonumber \\
&= \frac{1}{L^2} \sum_{p}^{}
e^{ipx}e^{-\hat{p}^2 t} \sum_{x'} e^{-ipx'} \phi_{x',0}.
\end{align}
Hence the final expression for the solution of Equation~(\ref{eq:he}) is
\begin{align}
\phi_{x,t} = \frac{1}{L^2} \sum_{p}^{} \sum_{x'} e^{-\hat{p}^2 t}
e^{ip(x-x')} \phi_{x',0}.
\end{align}

\subsection{Continuum smearing radius in $D$ dimensions}

Doing the same derivation in the continuum for $D$ dimensions one would get
\begin{align}
    \phi(x,t) &= \frac{1}{(2\pi)^{D}} \int d^{D}x' \int d^{D}p\, e^{-p^2 t}
    e^{ip(x-x')} \phi(x',0) \nonumber \\
              &\equiv \int d^{D}x'\, K_{t}(x-x') \phi(x',0),
      \label{eq:smoothing}
\end{align}
where we have defined the smearing kernel
\begin{align}
    K_{t}(z) = \int \frac{d^{D}p}{(2\pi)^{D}} e^{-p^2z}e^{ipz} =
    \frac{e^{-z^2 / 4t}}{(4\pi t)^{D / 2}}.
\end{align}
Analogously to the Yang--Mills gradient flow \cite{Luscher:2010iy},
Equation~(\ref{eq:smoothing}) shows that the heat equation is a smoothing operation
with mean-square radius
\begin{align}
    R_{D}(t) \equiv \sqrt{\int d^{D}z\; z^2 K_{t}(z)} = \sqrt{2Dt}.
\end{align}
In this work, the flow time $t$ for the computation of observables in smeared
configurations was tuned so that $R_{2}(t) = \xi$.

\newpage
\section{Supplementary plots and tables}
\label{sec:supplementary-tables-plots}

\begin{figure*}[!t]
    \centering
    \includegraphics[width=0.49\linewidth]{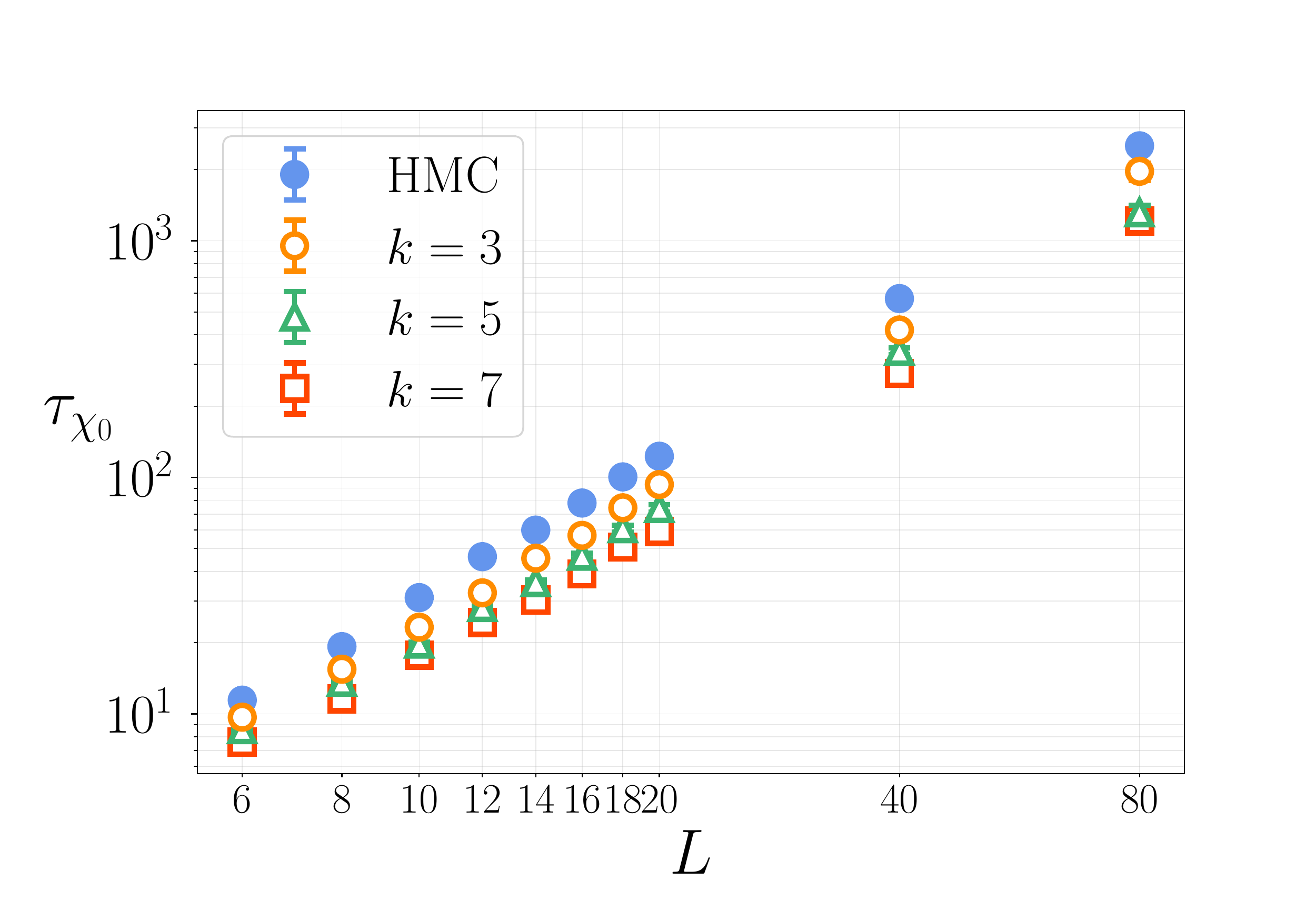}
    \includegraphics[width=0.49\linewidth]{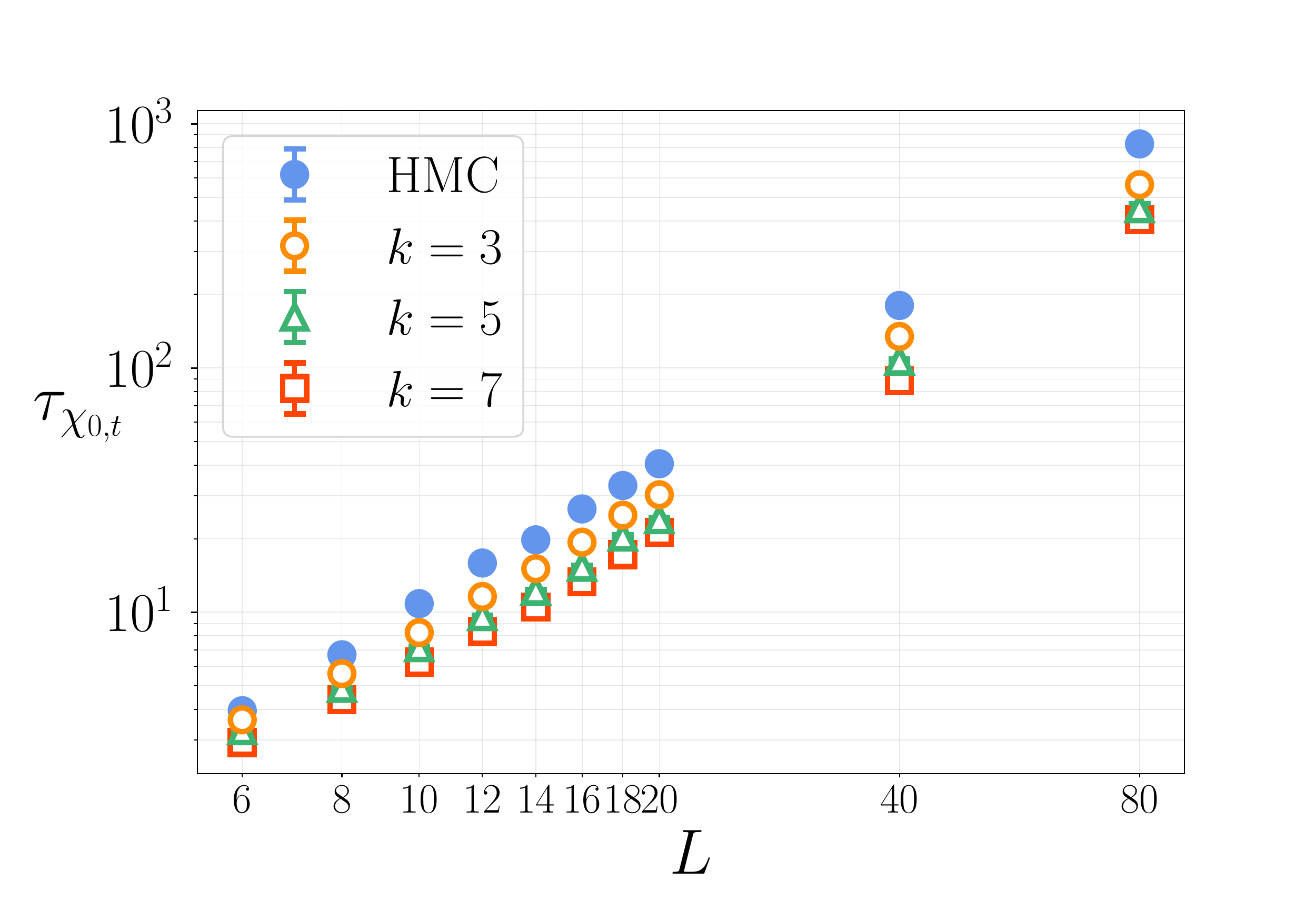}
    \caption{Scaling of the autocorrelation time of the unflowed (left) and
        flowed (right) one-point susceptibility towards the continuum for HMC
        (filled, blue circles) and FHMC with kernel sizes $k=3,5,7$ (open
        circles, triangles and squares).} 
    \label{fig:tauchifixedarch}
\end{figure*}

Reference values from the simulations of HMC and FHMC with $k=3$ and $N_{l}=1$
can be found in Tables~\ref{tab:hmcresults} and \ref{tb:flowhmc-k3},
respectively. Autocorrelation times for the unflowed and flowed one-point
susceptibilities, $\chi_{0}$ and $\chi_{0,t}$, are displayed in
Figure~\ref{fig:tauchifixedarch}, showing a similar behaviour to the
autocorrelation time of the magnetization showed in Figure~\ref{fig:Mfixedarch}.

\begin{table*}[t!]
    \begin{tabular}{r r r r r r r r r r r r}
        $L$ & $\tau_M$ & $|M|$ & $\tau_{|M|}$ & $\chi_0$ & $\tau_{\chi_0}$ & $\chi_{0,t}$ &
        $\tau_{\chi_{0,t}}$ & $\text{\# confs.}$ & $\text{\# steps}$ & $\text{acc.}$ \\
        \hline
        \hline
        6 &  11.439(94) & 0.27545(12) & 4.218(22) & 0.604082(45) & 1.3407(41) & 0.168701(79) & 3.953(20) & $20 \times 10^6$ & 5 & 0.91 \\
        8 &  19.26(20) & 0.26282(15) & 7.087(47) & 0.636218(46) & 2.0910(79) & 0.147630(92) & 6.696(43) & $20 \times 10^6$ & 6 & 0.91 \\
        10 & 31.03(40) & 0.25907(19) & 11.518(95) & 0.663980(49) & 3.284(15) & 0.13964(11) & 10.845(87) & $20 \times 10^6$ & 6 & 0.88 \\
        12 & 46.21(73) & 0.25362(22) & 16.89(17) & 0.684141(51) & 4.583(25) & 0.13247(13) & 15.90(15) & $20 \times 10^6$ & 6 & 0.86 \\
        14 & 59.8(11) & 0.24714(24) & 21.09(23) & 0.699527(49) & 5.367(31) & 0.12576(13) & 19.78(21) & $20 \times 10^6$ & 8 & 0.91 \\
        16 & 77.9(15) & 0.24368(28) & 28.36(35) & 0.713385(51) & 6.930(45) & 0.12175(15) & 26.48(32) & $20 \times 10^6$ & 8 & 0.89 \\
        18 & 100.4(16) & 0.24436(22) & 35.02(35) & 0.727030(37) & 8.493(44) & 0.12087(12) & 33.03(32) & $40 \times 10^6$ & 10 & 0.92 \\
        20 & 122.8(22) & 0.23818(23) & 43.30(48) & 0.735209(36) & 10.011(56) & 0.11562(12) & 40.55(44) & $40 \times 10^6$ & 10 & 0.91 \\
        40 & 570(21) & 0.22642(46) & 191.7(42) & 0.792768(41) & 36.88(38) & 0.10217(22) & 180.3(39) & $40 \times 10^6$ & 14 & 0.91 \\
        80 & 2518(130) & 0.20803(65) & 888(29) & 0.829350(31) & 134.3(18) & 0.08642(29) & 827(26) & $80 \times 10^6$ & 18 & 0.88 \\
    \end{tabular}
    \caption{HMC results.}
    \label{tab:hmcresults}
\end{table*}

\begin{table*}[t!]
\begin{tabular}{r r r r r r r r r r r r}
    $L$ &  $\tau_M$ & $|M|$ & $\tau_{|M|}$ & $\chi_0$ & $\tau_{\chi_0}$ & $\chi_{0,t}$ &
    $\tau_{\chi_{0,t}}$ & $\text{\# confs.}$ & $\text{\# steps}$ & $\text{acc.}$ \\
    \hline
    \hline
    6 &  9.69(14) & 0.27561(23) & 3.686(34) & 0.604119(96) & 1.5298(95) & 0.16877(15) & 3.627(34) & $5 \times 10^6$ & 8 & 0.98 \\
    8 &  15.46(28) & 0.26278(27) & 5.739(65) & 0.636246(94) & 2.225(16) & 0.14760(17) & 5.605(63) & $5 \times 10^6$ & 8 & 0.97 \\
    10 &  23.26(50) & 0.25913(32) & 8.53(12) & 0.663968(95) & 3.062(26) & 0.13964(19) & 8.25(11) & $5 \times 10^6$ & 8 & 0.95 \\
    12 &  32.50(82) & 0.25402(38) & 12.14(20) & 0.684240(96) & 4.047(39) & 0.13271(22) & 11.60(18) & $5 \times 10^6$ & 8 & 0.94 \\
    14 &  45.5(13) & 0.24735(42) & 15.84(29) & 0.699595(94) & 4.907(52) & 0.12593(23) & 15.07(27) & $5 \times 10^6$ & 10 & 0.95 \\
    16 &  56.9(18) & 0.24376(47) & 20.37(41) & 0.713427(94) & 5.972(69) & 0.12176(25) & 19.35(38) & $5 \times 10^6$ & 10 & 0.94 \\
    18 &  74.4(27) & 0.24472(53) & 26.40(60) & 0.727035(96) & 7.389(95) & 0.12098(28) & 24.97(56) & $5 \times 10^6$ & 10 & 0.93 \\
    20 &  93.2(38) & 0.23807(57) & 32.28(81) & 0.735258(95) & 8.47(12) & 0.11566(30) & 30.27(74) & $5 \times 10^6$ & 10 & 0.92 \\
    40 &  420(16) & 0.22596(48) & 143.5(33) & 0.792708(43) & 28.99(32) & 0.10204(23) & 134.6(30) & $30 \times 10^6$ & 15 & 0.89 \\
    80 &  1965(165) & 0.2077(11) & 611(30) & 0.829372(51) & 97.3(21) & 0.08638(46) & 563(27) & $20 \times 10^6$ & 25 & 0.87 \\
\end{tabular}
    \caption{FHMC results with a network with $N_{l} = 1$ and $k=3$.}
\label{tb:flowhmc-k3}
\end{table*}

\clearpage

\bibliography{biblio.bib}

\begin{thebibliography}{43}%
\makeatletter
\providecommand \@ifxundefined [1]{%
 \@ifx{#1\undefined}
}%
\providecommand \@ifnum [1]{%
 \ifnum #1\expandafter \@firstoftwo
 \else \expandafter \@secondoftwo
 \fi
}%
\providecommand \@ifx [1]{%
 \ifx #1\expandafter \@firstoftwo
 \else \expandafter \@secondoftwo
 \fi
}%
\providecommand \natexlab [1]{#1}%
\providecommand \enquote  [1]{``#1''}%
\providecommand \bibnamefont  [1]{#1}%
\providecommand \bibfnamefont [1]{#1}%
\providecommand \citenamefont [1]{#1}%
\providecommand \href@noop [0]{\@secondoftwo}%
\providecommand \href [0]{\begingroup \@sanitize@url \@href}%
\providecommand \@href[1]{\@@startlink{#1}\@@href}%
\providecommand \@@href[1]{\endgroup#1\@@endlink}%
\providecommand \@sanitize@url [0]{\catcode `\\12\catcode `\$12\catcode
  `\&12\catcode `\#12\catcode `\^12\catcode `\_12\catcode `\%12\relax}%
\providecommand \@@startlink[1]{}%
\providecommand \@@endlink[0]{}%
\providecommand \url  [0]{\begingroup\@sanitize@url \@url }%
\providecommand \@url [1]{\endgroup\@href {#1}{\urlprefix }}%
\providecommand \urlprefix  [0]{URL }%
\providecommand \Eprint [0]{\href }%
\providecommand \doibase [0]{https://doi.org/}%
\providecommand \selectlanguage [0]{\@gobble}%
\providecommand \bibinfo  [0]{\@secondoftwo}%
\providecommand \bibfield  [0]{\@secondoftwo}%
\providecommand \translation [1]{[#1]}%
\providecommand \BibitemOpen [0]{}%
\providecommand \bibitemStop [0]{}%
\providecommand \bibitemNoStop [0]{.\EOS\space}%
\providecommand \EOS [0]{\spacefactor3000\relax}%
\providecommand \BibitemShut  [1]{\csname bibitem#1\endcsname}%
\let\auto@bib@innerbib\@empty
%</preamble>
\bibitem [{\citenamefont {Campostrini}\ \emph {et~al.}(1992)\citenamefont
  {Campostrini}, \citenamefont {Rossi},\ and\ \citenamefont
  {Vicari}}]{Campostrini1992}%
  \BibitemOpen
  \bibfield  {author} {\bibinfo {author} {\bibfnamefont {M.}~\bibnamefont
  {Campostrini}}, \bibinfo {author} {\bibfnamefont {P.}~\bibnamefont {Rossi}},\
  and\ \bibinfo {author} {\bibfnamefont {E.}~\bibnamefont {Vicari}},\
  }\bibfield  {title} {\bibinfo {title} {{Monte Carlo} simulation of
  {$\mathrm{CP}^{N-1}$} models},\ }\href
  {https://doi.org/10.1103/PhysRevD.46.2647} {\bibfield  {journal} {\bibinfo
  {journal} {Phys. Rev. D}\ }\textbf {\bibinfo {volume} {46}},\ \bibinfo
  {pages} {2647} (\bibinfo {year} {1992})}\BibitemShut {NoStop}%
\bibitem [{\citenamefont {Vicari}(1993)}]{Vicari1993}%
  \BibitemOpen
  \bibfield  {author} {\bibinfo {author} {\bibfnamefont {E.}~\bibnamefont
  {Vicari}},\ }\bibfield  {title} {\bibinfo {title} {Monte carlo simulation of
  lattice {$\mathbb{CP}^{N-1}$} models at large {$N$}},\ }\href
  {https://doi.org/10.1016/0370-2693(93)91517-Q} {\bibfield  {journal}
  {\bibinfo  {journal} {Phys. Lett. B}\ }\textbf {\bibinfo {volume} {309}},\
  \bibinfo {pages} {139} (\bibinfo {year} {1993})},\ \Eprint
  {https://arxiv.org/abs/9209025} {arXiv:9209025 [hep-lat]} \BibitemShut
  {NoStop}%
\bibitem [{\citenamefont {Del~Debbio}\ \emph {et~al.}(2004)\citenamefont
  {Del~Debbio}, \citenamefont {Manca},\ and\ \citenamefont
  {Vicari}}]{DelDebbio:2004xh}%
  \BibitemOpen
  \bibfield  {author} {\bibinfo {author} {\bibfnamefont {L.}~\bibnamefont
  {Del~Debbio}}, \bibinfo {author} {\bibfnamefont {G.~M.}\ \bibnamefont
  {Manca}},\ and\ \bibinfo {author} {\bibfnamefont {E.}~\bibnamefont
  {Vicari}},\ }\bibfield  {title} {\bibinfo {title} {{Critical slowing down of
  topological modes}},\ }\href {https://doi.org/10.1016/j.physletb.2004.05.038}
  {\bibfield  {journal} {\bibinfo  {journal} {Phys. Lett. B}\ }\textbf
  {\bibinfo {volume} {594}},\ \bibinfo {pages} {315} (\bibinfo {year}
  {2004})},\ \Eprint {https://arxiv.org/abs/hep-lat/0403001}
  {arXiv:hep-lat/0403001} \BibitemShut {NoStop}%
\bibitem [{\citenamefont {Engel}\ and\ \citenamefont
  {Schaefer}(2011)}]{Schaefer2011}%
  \BibitemOpen
  \bibfield  {author} {\bibinfo {author} {\bibfnamefont {G.~P.}\ \bibnamefont
  {Engel}}\ and\ \bibinfo {author} {\bibfnamefont {S.}~\bibnamefont
  {Schaefer}},\ }\bibfield  {title} {\bibinfo {title} {{Testing trivializing
  maps in the Hybrid Monte Carlo algorithm}},\ }\href
  {https://doi.org/10.1016/j.cpc.2011.05.004} {\bibfield  {journal} {\bibinfo
  {journal} {Comput. Phys. Commun.}\ }\textbf {\bibinfo {volume} {182}},\
  \bibinfo {pages} {2107} (\bibinfo {year} {2011})},\ \Eprint
  {https://arxiv.org/abs/1102.1852} {arXiv:1102.1852 [hep-lat]} \BibitemShut
  {NoStop}%
\bibitem [{\citenamefont {Flynn}\ \emph {et~al.}(2015)\citenamefont {Flynn},
  \citenamefont {J\"{u}ttner}, \citenamefont {Lawson},\ and\ \citenamefont
  {Sanfilippo}}]{Flynn2015}%
  \BibitemOpen
  \bibfield  {author} {\bibinfo {author} {\bibfnamefont {J.}~\bibnamefont
  {Flynn}}, \bibinfo {author} {\bibfnamefont {A.}~\bibnamefont {J\"{u}ttner}},
  \bibinfo {author} {\bibfnamefont {A.}~\bibnamefont {Lawson}},\ and\ \bibinfo
  {author} {\bibfnamefont {F.}~\bibnamefont {Sanfilippo}},\ }\href@noop {}
  {\bibinfo {title} {Precision study of critical slowing down in lattice
  simulations of the {$\mathbb{CP}^{N-1}$} model}} (\bibinfo {year} {2015}),\
  \Eprint {https://arxiv.org/abs/1504.06292} {arXiv:1504.06292 [hep-lat]}
  \BibitemShut {NoStop}%
\bibitem [{\citenamefont {Bonati}\ and\ \citenamefont
  {D'Elia}(2018)}]{Bonati2018a}%
  \BibitemOpen
  \bibfield  {author} {\bibinfo {author} {\bibfnamefont {C.}~\bibnamefont
  {Bonati}}\ and\ \bibinfo {author} {\bibfnamefont {M.}~\bibnamefont
  {D'Elia}},\ }\bibfield  {title} {\bibinfo {title} {Topological critical
  slowing down: variations on a toy model},\ }\href
  {https://doi.org/10.1103/physreve.98.013308} {\bibfield  {journal} {\bibinfo
  {journal} {Phys. Rev. E}\ }\textbf {\bibinfo {volume} {98}},\ \bibinfo
  {pages} {013308} (\bibinfo {year} {2018})},\ \Eprint
  {https://arxiv.org/abs/1709.10034} {arXiv:1709.10034 [hep-lat]} \BibitemShut
  {NoStop}%
\bibitem [{\citenamefont {Del~Debbio}\ \emph {et~al.}(2002)\citenamefont
  {Del~Debbio}, \citenamefont {Panagopoulos}, \citenamefont {Rossi},\ and\
  \citenamefont {Vicari}}]{DelDebbio2002}%
  \BibitemOpen
  \bibfield  {author} {\bibinfo {author} {\bibfnamefont {L.}~\bibnamefont
  {Del~Debbio}}, \bibinfo {author} {\bibfnamefont {H.}~\bibnamefont
  {Panagopoulos}}, \bibinfo {author} {\bibfnamefont {P.}~\bibnamefont
  {Rossi}},\ and\ \bibinfo {author} {\bibfnamefont {E.}~\bibnamefont
  {Vicari}},\ }\bibfield  {title} {\bibinfo {title} {{Spectrum of confining
  strings in SU(N) gauge theories}},\ }\href
  {https://doi.org/10.1088/1126-6708/2002/01/009} {\bibfield  {journal}
  {\bibinfo  {journal} {JHEP}\ }\textbf {\bibinfo {volume} {01}},\ \bibinfo
  {pages} {009}},\ \Eprint {https://arxiv.org/abs/hep-th/0111090}
  {arXiv:hep-th/0111090} \BibitemShut {NoStop}%
\bibitem [{\citenamefont {Alles}\ \emph {et~al.}(1996)\citenamefont {Alles},
  \citenamefont {Boyd}, \citenamefont {D'Elia}, \citenamefont {Di~Giacomo},\
  and\ \citenamefont {Vicari}}]{Alles1996}%
  \BibitemOpen
  \bibfield  {author} {\bibinfo {author} {\bibfnamefont {B.}~\bibnamefont
  {Alles}}, \bibinfo {author} {\bibfnamefont {G.}~\bibnamefont {Boyd}},
  \bibinfo {author} {\bibfnamefont {M.}~\bibnamefont {D'Elia}}, \bibinfo
  {author} {\bibfnamefont {A.}~\bibnamefont {Di~Giacomo}},\ and\ \bibinfo
  {author} {\bibfnamefont {E.}~\bibnamefont {Vicari}},\ }\bibfield  {title}
  {\bibinfo {title} {{Hybrid Monte Carlo and topological modes of full QCD}},\
  }\href {https://doi.org/10.1016/S0370-2693(96)01247-6} {\bibfield  {journal}
  {\bibinfo  {journal} {Phys. Lett. B}\ }\textbf {\bibinfo {volume} {389}},\
  \bibinfo {pages} {107} (\bibinfo {year} {1996})},\ \Eprint
  {https://arxiv.org/abs/hep-lat/9607049} {arXiv:hep-lat/9607049} \BibitemShut
  {NoStop}%
\bibitem [{\citenamefont {L\"uscher}(2010)}]{Luscher:2010iy}%
  \BibitemOpen
  \bibfield  {author} {\bibinfo {author} {\bibfnamefont {M.}~\bibnamefont
  {L\"uscher}},\ }\bibfield  {title} {\bibinfo {title} {{Properties and uses of
  the Wilson flow in lattice QCD}},\ }\href
  {https://doi.org/10.1007/JHEP08(2010)071} {\bibfield  {journal} {\bibinfo
  {journal} {JHEP}\ }\textbf {\bibinfo {volume} {08}},\ \bibinfo {pages}
  {071}},\ \bibinfo {note} {[Erratum: JHEP 03, 092 (2014)]},\ \Eprint
  {https://arxiv.org/abs/1006.4518} {arXiv:1006.4518 [hep-lat]} \BibitemShut
  {NoStop}%
\bibitem [{\citenamefont {Schaefer}\ \emph {et~al.}(2011)\citenamefont
  {Schaefer}, \citenamefont {Sommer},\ and\ \citenamefont
  {Virotta}}]{Virotta2010}%
  \BibitemOpen
  \bibfield  {author} {\bibinfo {author} {\bibfnamefont {S.}~\bibnamefont
  {Schaefer}}, \bibinfo {author} {\bibfnamefont {R.}~\bibnamefont {Sommer}},\
  and\ \bibinfo {author} {\bibfnamefont {F.}~\bibnamefont {Virotta}} (\bibinfo
  {collaboration} {ALPHA}),\ }\bibfield  {title} {\bibinfo {title} {{Critical
  slowing down and error analysis in lattice QCD simulations}},\ }\href
  {https://doi.org/10.1016/j.nuclphysb.2010.11.020} {\bibfield  {journal}
  {\bibinfo  {journal} {Nucl. Phys. B}\ }\textbf {\bibinfo {volume} {845}},\
  \bibinfo {pages} {93} (\bibinfo {year} {2011})},\ \Eprint
  {https://arxiv.org/abs/1009.5228} {arXiv:1009.5228 [hep-lat]} \BibitemShut
  {NoStop}%
\bibitem [{\citenamefont {Luscher}(2010)}]{LuscherTrivializingMaps}%
  \BibitemOpen
  \bibfield  {author} {\bibinfo {author} {\bibfnamefont {M.}~\bibnamefont
  {Luscher}},\ }\bibfield  {title} {\bibinfo {title} {{Trivializing maps, the
  Wilson flow and the HMC algorithm}},\ }\href
  {https://doi.org/10.1007/s00220-009-0953-7} {\bibfield  {journal} {\bibinfo
  {journal} {Commun. Math. Phys.}\ }\textbf {\bibinfo {volume} {293}},\
  \bibinfo {pages} {899} (\bibinfo {year} {2010})},\ \Eprint
  {https://arxiv.org/abs/0907.5491} {arXiv:0907.5491 [hep-lat]} \BibitemShut
  {NoStop}%
\bibitem [{\citenamefont {Albergo}\ \emph {et~al.}(2019)\citenamefont
  {Albergo}, \citenamefont {Kanwar},\ and\ \citenamefont
  {Shanahan}}]{Albergo2019}%
  \BibitemOpen
  \bibfield  {author} {\bibinfo {author} {\bibfnamefont {M.~S.}\ \bibnamefont
  {Albergo}}, \bibinfo {author} {\bibfnamefont {G.}~\bibnamefont {Kanwar}},\
  and\ \bibinfo {author} {\bibfnamefont {P.~E.}\ \bibnamefont {Shanahan}},\
  }\bibfield  {title} {\bibinfo {title} {Flow-based generative models for
  {Markov chain Monte Carlo} in lattice field theory},\ }\href
  {https://doi.org/10.1103/physrevd.100.034515} {\bibfield  {journal} {\bibinfo
   {journal} {Phys. Rev. D}\ }\textbf {\bibinfo {volume} {100}},\ \bibinfo
  {pages} {034515} (\bibinfo {year} {2019})},\ \Eprint
  {https://arxiv.org/abs/1904.12072} {arXiv:1904.12072 [hep-lat]} \BibitemShut
  {NoStop}%
\bibitem [{\citenamefont {Kanwar}\ \emph {et~al.}(2020)\citenamefont {Kanwar},
  \citenamefont {Albergo}, \citenamefont {Boyda}, \citenamefont {Cranmer},
  \citenamefont {Hackett}, \citenamefont {Racani\`ere}, \citenamefont
  {Rezende},\ and\ \citenamefont {Shanahan}}]{Kanwar2020}%
  \BibitemOpen
  \bibfield  {author} {\bibinfo {author} {\bibfnamefont {G.}~\bibnamefont
  {Kanwar}}, \bibinfo {author} {\bibfnamefont {M.~S.}\ \bibnamefont {Albergo}},
  \bibinfo {author} {\bibfnamefont {D.}~\bibnamefont {Boyda}}, \bibinfo
  {author} {\bibfnamefont {K.}~\bibnamefont {Cranmer}}, \bibinfo {author}
  {\bibfnamefont {D.~C.}\ \bibnamefont {Hackett}}, \bibinfo {author}
  {\bibfnamefont {S.}~\bibnamefont {Racani\`ere}}, \bibinfo {author}
  {\bibfnamefont {D.~J.}\ \bibnamefont {Rezende}},\ and\ \bibinfo {author}
  {\bibfnamefont {P.~E.}\ \bibnamefont {Shanahan}},\ }\bibfield  {title}
  {\bibinfo {title} {{Equivariant flow-based sampling for lattice gauge
  theory}},\ }\href {https://doi.org/10.1103/PhysRevLett.125.121601} {\bibfield
   {journal} {\bibinfo  {journal} {Phys. Rev. Lett.}\ }\textbf {\bibinfo
  {volume} {125}},\ \bibinfo {pages} {121601} (\bibinfo {year} {2020})},\
  \Eprint {https://arxiv.org/abs/2003.06413} {arXiv:2003.06413 [hep-lat]}
  \BibitemShut {NoStop}%
\bibitem [{\citenamefont {Nicoli}\ \emph {et~al.}(2021)\citenamefont {Nicoli},
  \citenamefont {Anders}, \citenamefont {Funcke}, \citenamefont {Hartung},
  \citenamefont {Jansen}, \citenamefont {Kessel}, \citenamefont {Nakajima},\
  and\ \citenamefont {Stornati}}]{Nicoli2020}%
  \BibitemOpen
  \bibfield  {author} {\bibinfo {author} {\bibfnamefont {K.~A.}\ \bibnamefont
  {Nicoli}}, \bibinfo {author} {\bibfnamefont {C.~J.}\ \bibnamefont {Anders}},
  \bibinfo {author} {\bibfnamefont {L.}~\bibnamefont {Funcke}}, \bibinfo
  {author} {\bibfnamefont {T.}~\bibnamefont {Hartung}}, \bibinfo {author}
  {\bibfnamefont {K.}~\bibnamefont {Jansen}}, \bibinfo {author} {\bibfnamefont
  {P.}~\bibnamefont {Kessel}}, \bibinfo {author} {\bibfnamefont
  {S.}~\bibnamefont {Nakajima}},\ and\ \bibinfo {author} {\bibfnamefont
  {P.}~\bibnamefont {Stornati}},\ }\bibfield  {title} {\bibinfo {title}
  {{Estimation of Thermodynamic Observables in Lattice Field Theories with Deep
  Generative Models}},\ }\href {https://doi.org/10.1103/PhysRevLett.126.032001}
  {\bibfield  {journal} {\bibinfo  {journal} {Phys. Rev. Lett.}\ }\textbf
  {\bibinfo {volume} {126}},\ \bibinfo {pages} {032001} (\bibinfo {year}
  {2021})},\ \Eprint {https://arxiv.org/abs/2007.07115} {arXiv:2007.07115
  [hep-lat]} \BibitemShut {NoStop}%
\bibitem [{\citenamefont {Boyda}\ \emph {et~al.}(2021)\citenamefont {Boyda},
  \citenamefont {Kanwar}, \citenamefont {Racani\`ere}, \citenamefont {Rezende},
  \citenamefont {Albergo}, \citenamefont {Cranmer}, \citenamefont {Hackett},\
  and\ \citenamefont {Shanahan}}]{Boyda2020}%
  \BibitemOpen
  \bibfield  {author} {\bibinfo {author} {\bibfnamefont {D.}~\bibnamefont
  {Boyda}}, \bibinfo {author} {\bibfnamefont {G.}~\bibnamefont {Kanwar}},
  \bibinfo {author} {\bibfnamefont {S.}~\bibnamefont {Racani\`ere}}, \bibinfo
  {author} {\bibfnamefont {D.~J.}\ \bibnamefont {Rezende}}, \bibinfo {author}
  {\bibfnamefont {M.~S.}\ \bibnamefont {Albergo}}, \bibinfo {author}
  {\bibfnamefont {K.}~\bibnamefont {Cranmer}}, \bibinfo {author} {\bibfnamefont
  {D.~C.}\ \bibnamefont {Hackett}},\ and\ \bibinfo {author} {\bibfnamefont
  {P.~E.}\ \bibnamefont {Shanahan}},\ }\bibfield  {title} {\bibinfo {title}
  {{Sampling using $SU(N)$ gauge equivariant flows}},\ }\href
  {https://doi.org/10.1103/PhysRevD.103.074504} {\bibfield  {journal} {\bibinfo
   {journal} {Phys. Rev. D}\ }\textbf {\bibinfo {volume} {103}},\ \bibinfo
  {pages} {074504} (\bibinfo {year} {2021})},\ \Eprint
  {https://arxiv.org/abs/2008.05456} {arXiv:2008.05456 [hep-lat]} \BibitemShut
  {NoStop}%
\bibitem [{\citenamefont {Albergo}\ \emph
  {et~al.}(2021{\natexlab{a}})\citenamefont {Albergo}, \citenamefont {Kanwar},
  \citenamefont {Racani\`ere}, \citenamefont {Rezende}, \citenamefont {Urban},
  \citenamefont {Boyda}, \citenamefont {Cranmer}, \citenamefont {Hackett},\
  and\ \citenamefont {Shanahan}}]{Albergo2021}%
  \BibitemOpen
  \bibfield  {author} {\bibinfo {author} {\bibfnamefont {M.~S.}\ \bibnamefont
  {Albergo}}, \bibinfo {author} {\bibfnamefont {G.}~\bibnamefont {Kanwar}},
  \bibinfo {author} {\bibfnamefont {S.}~\bibnamefont {Racani\`ere}}, \bibinfo
  {author} {\bibfnamefont {D.~J.}\ \bibnamefont {Rezende}}, \bibinfo {author}
  {\bibfnamefont {J.~M.}\ \bibnamefont {Urban}}, \bibinfo {author}
  {\bibfnamefont {D.}~\bibnamefont {Boyda}}, \bibinfo {author} {\bibfnamefont
  {K.}~\bibnamefont {Cranmer}}, \bibinfo {author} {\bibfnamefont {D.~C.}\
  \bibnamefont {Hackett}},\ and\ \bibinfo {author} {\bibfnamefont {P.~E.}\
  \bibnamefont {Shanahan}},\ }\bibfield  {title} {\bibinfo {title} {{Flow-based
  sampling for fermionic lattice field theories}},\ }\href
  {https://doi.org/10.1103/PhysRevD.104.114507} {\bibfield  {journal} {\bibinfo
   {journal} {Phys. Rev. D}\ }\textbf {\bibinfo {volume} {104}},\ \bibinfo
  {pages} {114507} (\bibinfo {year} {2021}{\natexlab{a}})},\ \Eprint
  {https://arxiv.org/abs/2106.05934} {arXiv:2106.05934 [hep-lat]} \BibitemShut
  {NoStop}%
\bibitem [{\citenamefont {Albergo}\ \emph {et~al.}(2022)\citenamefont
  {Albergo}, \citenamefont {Boyda}, \citenamefont {Cranmer}, \citenamefont
  {Hackett}, \citenamefont {Kanwar}, \citenamefont {Racani\`ere}, \citenamefont
  {Rezende}, \citenamefont {Romero-L\'opez}, \citenamefont {Shanahan},\ and\
  \citenamefont {Urban}}]{Albergo2022}%
  \BibitemOpen
  \bibfield  {author} {\bibinfo {author} {\bibfnamefont {M.~S.}\ \bibnamefont
  {Albergo}}, \bibinfo {author} {\bibfnamefont {D.}~\bibnamefont {Boyda}},
  \bibinfo {author} {\bibfnamefont {K.}~\bibnamefont {Cranmer}}, \bibinfo
  {author} {\bibfnamefont {D.~C.}\ \bibnamefont {Hackett}}, \bibinfo {author}
  {\bibfnamefont {G.}~\bibnamefont {Kanwar}}, \bibinfo {author} {\bibfnamefont
  {S.}~\bibnamefont {Racani\`ere}}, \bibinfo {author} {\bibfnamefont {D.~J.}\
  \bibnamefont {Rezende}}, \bibinfo {author} {\bibfnamefont {F.}~\bibnamefont
  {Romero-L\'opez}}, \bibinfo {author} {\bibfnamefont {P.~E.}\ \bibnamefont
  {Shanahan}},\ and\ \bibinfo {author} {\bibfnamefont {J.~M.}\ \bibnamefont
  {Urban}},\ }\bibfield  {title} {\bibinfo {title} {{Flow-based sampling in the
  lattice Schwinger model at criticality}},\ }\href
  {https://doi.org/10.1103/PhysRevD.106.014514} {\bibfield  {journal} {\bibinfo
   {journal} {Phys. Rev. D}\ }\textbf {\bibinfo {volume} {106}},\ \bibinfo
  {pages} {014514} (\bibinfo {year} {2022})},\ \Eprint
  {https://arxiv.org/abs/2202.11712} {arXiv:2202.11712 [hep-lat]} \BibitemShut
  {NoStop}%
\bibitem [{\citenamefont {Abbott}\ \emph
  {et~al.}(2022{\natexlab{a}})\citenamefont {Abbott} \emph
  {et~al.}}]{Abbott2022}%
  \BibitemOpen
  \bibfield  {author} {\bibinfo {author} {\bibfnamefont {R.}~\bibnamefont
  {Abbott}} \emph {et~al.},\ }\bibfield  {title} {\bibinfo {title}
  {{Gauge-equivariant flow models for sampling in lattice field theories with
  pseudofermions}},\ }\href {https://doi.org/10.1103/PhysRevD.106.074506}
  {\bibfield  {journal} {\bibinfo  {journal} {Phys. Rev. D}\ }\textbf {\bibinfo
  {volume} {106}},\ \bibinfo {pages} {074506} (\bibinfo {year}
  {2022}{\natexlab{a}})},\ \Eprint {https://arxiv.org/abs/2207.08945}
  {arXiv:2207.08945 [hep-lat]} \BibitemShut {NoStop}%
\bibitem [{\citenamefont {Tabak}\ and\ \citenamefont
  {Vanden-Eijnden}(2010)}]{Tabak2010}%
  \BibitemOpen
  \bibfield  {author} {\bibinfo {author} {\bibfnamefont {E.~G.}\ \bibnamefont
  {Tabak}}\ and\ \bibinfo {author} {\bibfnamefont {E.}~\bibnamefont
  {Vanden-Eijnden}},\ }\bibfield  {title} {\bibinfo {title} {Density estimation
  by dual ascent of the log-likelihood},\ }\bibfield  {journal} {\bibinfo
  {journal} {Communications in Mathematical Sciences}\ }\textbf {\bibinfo
  {volume} {8}},\ \href {https://doi.org/10.4310/cms.2010.v8.n1.a11}
  {10.4310/cms.2010.v8.n1.a11} (\bibinfo {year} {2010})\BibitemShut {NoStop}%
\bibitem [{\citenamefont {Tabak}\ and\ \citenamefont
  {Turner}(2012)}]{Tabak2012}%
  \BibitemOpen
  \bibfield  {author} {\bibinfo {author} {\bibfnamefont {E.~G.}\ \bibnamefont
  {Tabak}}\ and\ \bibinfo {author} {\bibfnamefont {C.~V.}\ \bibnamefont
  {Turner}},\ }\bibfield  {title} {\bibinfo {title} {A family of nonparametric
  density estimation algorithms},\ }\href {https://doi.org/10.1002/cpa.21423}
  {\bibfield  {journal} {\bibinfo  {journal} {Commun. Pure Appl. Math.}\
  }\textbf {\bibinfo {volume} {66}},\ \bibinfo {pages} {145} (\bibinfo {year}
  {2012})}\BibitemShut {NoStop}%
\bibitem [{\citenamefont {Rezende}\ and\ \citenamefont
  {Mohamed}(2015)}]{Rezende2015}%
  \BibitemOpen
  \bibfield  {author} {\bibinfo {author} {\bibfnamefont {D.~J.}\ \bibnamefont
  {Rezende}}\ and\ \bibinfo {author} {\bibfnamefont {S.}~\bibnamefont
  {Mohamed}},\ }\href {https://doi.org/10.48550/ARXIV.1505.05770} {\bibinfo
  {title} {Variational inference with normalizing flows}} (\bibinfo {year}
  {2015}),\ \Eprint {https://arxiv.org/abs/1505.05770} {arXiv:1505.05770
  [stat.ML]} \BibitemShut {NoStop}%
\bibitem [{\citenamefont {Del~Debbio}\ \emph {et~al.}(2021)\citenamefont
  {Del~Debbio}, \citenamefont {Marsh~Rossney},\ and\ \citenamefont
  {Wilson}}]{DelDebbio:2021qwf}%
  \BibitemOpen
  \bibfield  {author} {\bibinfo {author} {\bibfnamefont {L.}~\bibnamefont
  {Del~Debbio}}, \bibinfo {author} {\bibfnamefont {J.}~\bibnamefont
  {Marsh~Rossney}},\ and\ \bibinfo {author} {\bibfnamefont {M.}~\bibnamefont
  {Wilson}},\ }\bibfield  {title} {\bibinfo {title} {{Efficient modeling of
  trivializing maps for lattice \ensuremath{\phi}4 theory using normalizing
  flows: A first look at scalability}},\ }\href
  {https://doi.org/10.1103/PhysRevD.104.094507} {\bibfield  {journal} {\bibinfo
   {journal} {Phys. Rev. D}\ }\textbf {\bibinfo {volume} {104}},\ \bibinfo
  {pages} {094507} (\bibinfo {year} {2021})},\ \Eprint
  {https://arxiv.org/abs/2105.12481} {arXiv:2105.12481 [hep-lat]} \BibitemShut
  {NoStop}%
\bibitem [{\citenamefont {Albandea}\ \emph {et~al.}(2022)\citenamefont
  {Albandea}, \citenamefont {Del~Debbio}, \citenamefont {Hern\'andez},
  \citenamefont {Kenway}, \citenamefont {Marsh~Rossney},\ and\ \citenamefont
  {Ramos}}]{Albandea:2022fky}%
  \BibitemOpen
  \bibfield  {author} {\bibinfo {author} {\bibfnamefont {D.}~\bibnamefont
  {Albandea}}, \bibinfo {author} {\bibfnamefont {L.}~\bibnamefont
  {Del~Debbio}}, \bibinfo {author} {\bibfnamefont {P.}~\bibnamefont
  {Hern\'andez}}, \bibinfo {author} {\bibfnamefont {R.}~\bibnamefont {Kenway}},
  \bibinfo {author} {\bibfnamefont {J.}~\bibnamefont {Marsh~Rossney}},\ and\
  \bibinfo {author} {\bibfnamefont {A.}~\bibnamefont {Ramos}},\ }\bibfield
  {title} {\bibinfo {title} {{Learning trivializing flows}},\ }in\ \href@noop
  {} {\emph {\bibinfo {booktitle} {{39th International Symposium on Lattice
  Field Theory}}}}\ (\bibinfo {year} {2022})\ \Eprint
  {https://arxiv.org/abs/2211.12806} {arXiv:2211.12806 [hep-lat]} \BibitemShut
  {NoStop}%
\bibitem [{\citenamefont {Dinh}\ \emph {et~al.}(2014)\citenamefont {Dinh},
  \citenamefont {Krueger},\ and\ \citenamefont {Bengio}}]{Dinh2014}%
  \BibitemOpen
  \bibfield  {author} {\bibinfo {author} {\bibfnamefont {L.}~\bibnamefont
  {Dinh}}, \bibinfo {author} {\bibfnamefont {D.}~\bibnamefont {Krueger}},\ and\
  \bibinfo {author} {\bibfnamefont {Y.}~\bibnamefont {Bengio}},\ }\href@noop {}
  {\bibinfo {title} {{NICE}: Non-linear independent components estimation}}
  (\bibinfo {year} {2014}),\ \Eprint {https://arxiv.org/abs/1410.8516}
  {arXiv:1410.8516 [cs.LG]} \BibitemShut {NoStop}%
\bibitem [{\citenamefont {Dinh}\ \emph {et~al.}(2016)\citenamefont {Dinh},
  \citenamefont {Sohl-Dickstein},\ and\ \citenamefont {Bengio}}]{Dinh2016}%
  \BibitemOpen
  \bibfield  {author} {\bibinfo {author} {\bibfnamefont {L.}~\bibnamefont
  {Dinh}}, \bibinfo {author} {\bibfnamefont {J.}~\bibnamefont
  {Sohl-Dickstein}},\ and\ \bibinfo {author} {\bibfnamefont {S.}~\bibnamefont
  {Bengio}},\ }\href@noop {} {\bibinfo {title} {Density estimation using {Real
  NVP}}} (\bibinfo {year} {2016}),\ \Eprint {https://arxiv.org/abs/1605.08803}
  {arXiv:1605.08803 [cs.LG]} \BibitemShut {NoStop}%
\bibitem [{\citenamefont {Kingma}\ and\ \citenamefont
  {Dhariwal}(2018)}]{Kingma2018}%
  \BibitemOpen
  \bibfield  {author} {\bibinfo {author} {\bibfnamefont {D.~P.}\ \bibnamefont
  {Kingma}}\ and\ \bibinfo {author} {\bibfnamefont {P.}~\bibnamefont
  {Dhariwal}},\ }\href@noop {} {\bibinfo {title} {Glow: Generative flow with
  invertible 1x1 convolutions}} (\bibinfo {year} {2018}),\ \Eprint
  {https://arxiv.org/abs/1807.03039} {arXiv:1807.03039 [stat.ML]} \BibitemShut
  {NoStop}%
\bibitem [{\citenamefont {Kullbach}\ and\ \citenamefont
  {Leibler}(1951)}]{Kullbach1951}%
  \BibitemOpen
  \bibfield  {author} {\bibinfo {author} {\bibfnamefont {S.}~\bibnamefont
  {Kullbach}}\ and\ \bibinfo {author} {\bibfnamefont {R.~A.}\ \bibnamefont
  {Leibler}},\ }\bibfield  {title} {\bibinfo {title} {On information and
  sufficiency},\ }\href {https://doi.org/10.1214/aoms/1177729694} {\bibfield
  {journal} {\bibinfo  {journal} {Ann. Math. Statist.}\ }\textbf {\bibinfo
  {volume} {22}},\ \bibinfo {pages} {79} (\bibinfo {year} {1951})}\BibitemShut
  {NoStop}%
\bibitem [{\citenamefont {Kingma}\ and\ \citenamefont {Ba}(2014)}]{Kingma2014}%
  \BibitemOpen
  \bibfield  {author} {\bibinfo {author} {\bibfnamefont {D.~P.}\ \bibnamefont
  {Kingma}}\ and\ \bibinfo {author} {\bibfnamefont {J.}~\bibnamefont {Ba}},\
  }\href@noop {} {\bibinfo {title} {Adam: A method for stochastic
  optimization}} (\bibinfo {year} {2014}),\ \Eprint
  {https://arxiv.org/abs/1412.6980} {arXiv:1412.6980 [cs.LG]} \BibitemShut
  {NoStop}%
\bibitem [{\citenamefont {Metropolis}\ \emph {et~al.}(1953)\citenamefont
  {Metropolis}, \citenamefont {Rosenbluth}, \citenamefont {Rosenbluth},
  \citenamefont {Teller},\ and\ \citenamefont {Teller}}]{Metropolis1953}%
  \BibitemOpen
  \bibfield  {author} {\bibinfo {author} {\bibfnamefont {N.}~\bibnamefont
  {Metropolis}}, \bibinfo {author} {\bibfnamefont {A.~W.}\ \bibnamefont
  {Rosenbluth}}, \bibinfo {author} {\bibfnamefont {M.~N.}\ \bibnamefont
  {Rosenbluth}}, \bibinfo {author} {\bibfnamefont {A.~H.}\ \bibnamefont
  {Teller}},\ and\ \bibinfo {author} {\bibfnamefont {E.}~\bibnamefont
  {Teller}},\ }\bibfield  {title} {\bibinfo {title} {Equation of state
  calculations by fast computing machines},\ }\href
  {https://doi.org/10.2172/4390578} {\bibfield  {journal} {\bibinfo  {journal}
  {J. Chem. Phys.}\ }\textbf {\bibinfo {volume} {21}},\ \bibinfo {pages} {1087}
  (\bibinfo {year} {1953})}\BibitemShut {NoStop}%
\bibitem [{\citenamefont {Hastings}(1970)}]{Hastings1970}%
  \BibitemOpen
  \bibfield  {author} {\bibinfo {author} {\bibfnamefont {W.~K.}\ \bibnamefont
  {Hastings}},\ }\bibfield  {title} {\bibinfo {title} {{Monte Carlo Sampling
  Methods Using Markov Chains and Their Applications}},\ }\href
  {https://doi.org/10.1093/biomet/57.1.97} {\bibfield  {journal} {\bibinfo
  {journal} {Biometrika}\ }\textbf {\bibinfo {volume} {57}},\ \bibinfo {pages}
  {97} (\bibinfo {year} {1970})}\BibitemShut {NoStop}%
\bibitem [{\citenamefont {Foreman}\ \emph {et~al.}(2022)\citenamefont
  {Foreman}, \citenamefont {Izubuchi}, \citenamefont {Jin}, \citenamefont
  {Jin}, \citenamefont {Osborn},\ and\ \citenamefont {Tomiya}}]{Foreman2022}%
  \BibitemOpen
  \bibfield  {author} {\bibinfo {author} {\bibfnamefont {S.}~\bibnamefont
  {Foreman}}, \bibinfo {author} {\bibfnamefont {T.}~\bibnamefont {Izubuchi}},
  \bibinfo {author} {\bibfnamefont {L.}~\bibnamefont {Jin}}, \bibinfo {author}
  {\bibfnamefont {X.-y.}\ \bibnamefont {Jin}}, \bibinfo {author} {\bibfnamefont
  {J.~C.}\ \bibnamefont {Osborn}},\ and\ \bibinfo {author} {\bibfnamefont
  {A.}~\bibnamefont {Tomiya}},\ }\bibfield  {title} {\bibinfo {title} {{HMC
  with Normalizing Flows}},\ }\href {https://doi.org/10.22323/1.396.0073}
  {\bibfield  {journal} {\bibinfo  {journal} {PoS}\ }\textbf {\bibinfo {volume}
  {LATTICE2021}},\ \bibinfo {pages} {073} (\bibinfo {year} {2022})}\BibitemShut
  {NoStop}%
\bibitem [{\citenamefont {Jin}(2022)}]{Jin2022}%
  \BibitemOpen
  \bibfield  {author} {\bibinfo {author} {\bibfnamefont {X.-y.}\ \bibnamefont
  {Jin}},\ }\bibfield  {title} {\bibinfo {title} {{Neural Network Field
  Transformation and Its Application in HMC}},\ }\href
  {https://doi.org/10.22323/1.396.0600} {\bibfield  {journal} {\bibinfo
  {journal} {PoS}\ }\textbf {\bibinfo {volume} {LATTICE2021}},\ \bibinfo
  {pages} {600} (\bibinfo {year} {2022})}\BibitemShut {NoStop}%
\bibitem [{\citenamefont {Bacchio}\ \emph {et~al.}(2022)\citenamefont
  {Bacchio}, \citenamefont {Kessel}, \citenamefont {Schaefer},\ and\
  \citenamefont {Vaitl}}]{Bacchio2022}%
  \BibitemOpen
  \bibfield  {author} {\bibinfo {author} {\bibfnamefont {S.}~\bibnamefont
  {Bacchio}}, \bibinfo {author} {\bibfnamefont {P.}~\bibnamefont {Kessel}},
  \bibinfo {author} {\bibfnamefont {S.}~\bibnamefont {Schaefer}},\ and\
  \bibinfo {author} {\bibfnamefont {L.}~\bibnamefont {Vaitl}},\ }\bibfield
  {title} {\bibinfo {title} {{Learning Trivializing Gradient Flows for Lattice
  Gauge Theories}},\ }\href@noop {} {\  (\bibinfo {year} {2022})},\ \Eprint
  {https://arxiv.org/abs/2212.08469} {arXiv:2212.08469 [hep-lat]} \BibitemShut
  {NoStop}%
\bibitem [{\citenamefont {Wolff}(2004)}]{Wolff:2003sm}%
  \BibitemOpen
  \bibfield  {author} {\bibinfo {author} {\bibfnamefont {U.}~\bibnamefont
  {Wolff}} (\bibinfo {collaboration} {ALPHA}),\ }\bibfield  {title} {\bibinfo
  {title} {{Monte Carlo errors with less errors}},\ }\href
  {https://doi.org/10.1016/S0010-4655(03)00467-3} {\bibfield  {journal}
  {\bibinfo  {journal} {Comput. Phys. Commun.}\ }\textbf {\bibinfo {volume}
  {156}},\ \bibinfo {pages} {143} (\bibinfo {year} {2004})},\ \bibinfo {note}
  {[Erratum: Comput.Phys.Commun. 176, 383 (2007)]},\ \Eprint
  {https://arxiv.org/abs/hep-lat/0306017} {arXiv:hep-lat/0306017} \BibitemShut
  {NoStop}%
\bibitem [{\citenamefont {Ramos}(2021)}]{Ramos2020}%
  \BibitemOpen
  \bibfield  {author} {\bibinfo {author} {\bibfnamefont {A.}~\bibnamefont
  {Ramos}},\ }\bibfield  {title} {\bibinfo {title} {{Automatic differentiation
  for error analysis}},\ }\href {https://doi.org/10.22323/1.392.0045}
  {\bibfield  {journal} {\bibinfo  {journal} {PoS}\ }\textbf {\bibinfo {volume}
  {TOOLS2020}},\ \bibinfo {pages} {045} (\bibinfo {year} {2021})},\ \Eprint
  {https://arxiv.org/abs/2012.11183} {arXiv:2012.11183 [hep-lat]} \BibitemShut
  {NoStop}%
\bibitem [{\citenamefont {Ramos}(2019)}]{Ramos:2018vgu}%
  \BibitemOpen
  \bibfield  {author} {\bibinfo {author} {\bibfnamefont {A.}~\bibnamefont
  {Ramos}},\ }\bibfield  {title} {\bibinfo {title} {{Automatic differentiation
  for error analysis of Monte Carlo data}},\ }\href
  {https://doi.org/10.1016/j.cpc.2018.12.020} {\bibfield  {journal} {\bibinfo
  {journal} {Comput. Phys. Commun.}\ }\textbf {\bibinfo {volume} {238}},\
  \bibinfo {pages} {19} (\bibinfo {year} {2019})},\ \Eprint
  {https://arxiv.org/abs/1809.01289} {arXiv:1809.01289 [hep-lat]} \BibitemShut
  {NoStop}%
\bibitem [{\citenamefont {Griewank}\ and\ \citenamefont
  {Walther}(2008)}]{Griewank2008}%
  \BibitemOpen
  \bibfield  {author} {\bibinfo {author} {\bibfnamefont {A.}~\bibnamefont
  {Griewank}}\ and\ \bibinfo {author} {\bibfnamefont {A.}~\bibnamefont
  {Walther}},\ }\href@noop {} {\emph {\bibinfo {title} {Evaluating Derivatives:
  Principles and Techniques of Algorithmic Differentiation}}},\ \bibinfo
  {edition} {2nd}\ ed.\ (\bibinfo  {publisher} {Society for Industrial and
  Applied Mathematics},\ \bibinfo {address} {USA},\ \bibinfo {year}
  {2008})\BibitemShut {NoStop}%
\bibitem [{\citenamefont {Albergo}\ \emph
  {et~al.}(2021{\natexlab{b}})\citenamefont {Albergo}, \citenamefont {Boyda},
  \citenamefont {Hackett}, \citenamefont {Kanwar}, \citenamefont {Cranmer},
  \citenamefont {Racani\`ere}, \citenamefont {Rezende},\ and\ \citenamefont
  {Shanahan}}]{Albergo:2021vyo}%
  \BibitemOpen
  \bibfield  {author} {\bibinfo {author} {\bibfnamefont {M.~S.}\ \bibnamefont
  {Albergo}}, \bibinfo {author} {\bibfnamefont {D.}~\bibnamefont {Boyda}},
  \bibinfo {author} {\bibfnamefont {D.~C.}\ \bibnamefont {Hackett}}, \bibinfo
  {author} {\bibfnamefont {G.}~\bibnamefont {Kanwar}}, \bibinfo {author}
  {\bibfnamefont {K.}~\bibnamefont {Cranmer}}, \bibinfo {author} {\bibfnamefont
  {S.}~\bibnamefont {Racani\`ere}}, \bibinfo {author} {\bibfnamefont {D.~J.}\
  \bibnamefont {Rezende}},\ and\ \bibinfo {author} {\bibfnamefont {P.~E.}\
  \bibnamefont {Shanahan}},\ }\bibfield  {title} {\bibinfo {title}
  {{Introduction to Normalizing Flows for Lattice Field Theory}},\ }\href@noop
  {} {\  (\bibinfo {year} {2021}{\natexlab{b}})},\ \Eprint
  {https://arxiv.org/abs/2101.08176} {arXiv:2101.08176 [hep-lat]} \BibitemShut
  {NoStop}%
\bibitem [{\citenamefont {Baulieu}\ and\ \citenamefont
  {Zwanziger}(2000)}]{Baulieu:1999wz}%
  \BibitemOpen
  \bibfield  {author} {\bibinfo {author} {\bibfnamefont {L.}~\bibnamefont
  {Baulieu}}\ and\ \bibinfo {author} {\bibfnamefont {D.}~\bibnamefont
  {Zwanziger}},\ }\bibfield  {title} {\bibinfo {title} {{QCD(4) from a
  five-dimensional point of view}},\ }\href
  {https://doi.org/10.1016/S0550-3213(00)00176-0} {\bibfield  {journal}
  {\bibinfo  {journal} {Nucl. Phys. B}\ }\textbf {\bibinfo {volume} {581}},\
  \bibinfo {pages} {604} (\bibinfo {year} {2000})},\ \Eprint
  {https://arxiv.org/abs/hep-th/9909006} {arXiv:hep-th/9909006} \BibitemShut
  {NoStop}%
\bibitem [{\citenamefont {Paszke}\ \emph {et~al.}(2019)\citenamefont {Paszke},
  \citenamefont {Gross}, \citenamefont {Massa}, \citenamefont {Lerer},
  \citenamefont {Bradbury}, \citenamefont {Chanan}, \citenamefont {Killeen},
  \citenamefont {Lin}, \citenamefont {Gimelshein}, \citenamefont {Antiga},
  \citenamefont {Desmaison}, \citenamefont {Kopf}, \citenamefont {Yang},
  \citenamefont {DeVito}, \citenamefont {Raison}, \citenamefont {Tejani},
  \citenamefont {Chilamkurthy}, \citenamefont {Steiner}, \citenamefont {Fang},
  \citenamefont {Bai},\ and\ \citenamefont {Chintala}}]{NEURIPS2019_bdbca288}%
  \BibitemOpen
  \bibfield  {author} {\bibinfo {author} {\bibfnamefont {A.}~\bibnamefont
  {Paszke}}, \bibinfo {author} {\bibfnamefont {S.}~\bibnamefont {Gross}},
  \bibinfo {author} {\bibfnamefont {F.}~\bibnamefont {Massa}}, \bibinfo
  {author} {\bibfnamefont {A.}~\bibnamefont {Lerer}}, \bibinfo {author}
  {\bibfnamefont {J.}~\bibnamefont {Bradbury}}, \bibinfo {author}
  {\bibfnamefont {G.}~\bibnamefont {Chanan}}, \bibinfo {author} {\bibfnamefont
  {T.}~\bibnamefont {Killeen}}, \bibinfo {author} {\bibfnamefont
  {Z.}~\bibnamefont {Lin}}, \bibinfo {author} {\bibfnamefont {N.}~\bibnamefont
  {Gimelshein}}, \bibinfo {author} {\bibfnamefont {L.}~\bibnamefont {Antiga}},
  \bibinfo {author} {\bibfnamefont {A.}~\bibnamefont {Desmaison}}, \bibinfo
  {author} {\bibfnamefont {A.}~\bibnamefont {Kopf}}, \bibinfo {author}
  {\bibfnamefont {E.}~\bibnamefont {Yang}}, \bibinfo {author} {\bibfnamefont
  {Z.}~\bibnamefont {DeVito}}, \bibinfo {author} {\bibfnamefont
  {M.}~\bibnamefont {Raison}}, \bibinfo {author} {\bibfnamefont
  {A.}~\bibnamefont {Tejani}}, \bibinfo {author} {\bibfnamefont
  {S.}~\bibnamefont {Chilamkurthy}}, \bibinfo {author} {\bibfnamefont
  {B.}~\bibnamefont {Steiner}}, \bibinfo {author} {\bibfnamefont
  {L.}~\bibnamefont {Fang}}, \bibinfo {author} {\bibfnamefont {J.}~\bibnamefont
  {Bai}},\ and\ \bibinfo {author} {\bibfnamefont {S.}~\bibnamefont
  {Chintala}},\ }\bibfield  {title} {\bibinfo {title} {Pytorch: An imperative
  style, high-performance deep learning library},\ }in\ \href
  {https://proceedings.neurips.cc/paper/2019/file/bdbca288fee7f92f2bfa9f7012727740-Paper.pdf}
  {\emph {\bibinfo {booktitle} {Advances in Neural Information Processing
  Systems}}},\ Vol.~\bibinfo {volume} {32},\ \bibinfo {editor} {edited by\
  \bibinfo {editor} {\bibfnamefont {H.}~\bibnamefont {Wallach}}, \bibinfo
  {editor} {\bibfnamefont {H.}~\bibnamefont {Larochelle}}, \bibinfo {editor}
  {\bibfnamefont {A.}~\bibnamefont {Beygelzimer}}, \bibinfo {editor}
  {\bibfnamefont {F.}~\bibnamefont {d\textquotesingle Alch\'{e}-Buc}}, \bibinfo
  {editor} {\bibfnamefont {E.}~\bibnamefont {Fox}},\ and\ \bibinfo {editor}
  {\bibfnamefont {R.}~\bibnamefont {Garnett}}}\ (\bibinfo  {publisher} {Curran
  Associates, Inc.},\ \bibinfo {year} {2019})\BibitemShut {NoStop}%
\bibitem [{\citenamefont {Madras}\ and\ \citenamefont
  {Sokal}(1988)}]{Madras1988}%
  \BibitemOpen
  \bibfield  {author} {\bibinfo {author} {\bibfnamefont {N.}~\bibnamefont
  {Madras}}\ and\ \bibinfo {author} {\bibfnamefont {A.~D.}\ \bibnamefont
  {Sokal}},\ }\bibfield  {title} {\bibinfo {title} {The pivot algorithm: A
  highly efficient monte carlo method for the self-avoiding walk},\ }\href
  {https://doi.org/10.1007/BF01022990} {\bibfield  {journal} {\bibinfo
  {journal} {Journal of Statistical Physics}\ }\textbf {\bibinfo {volume}
  {50}},\ \bibinfo {pages} {109} (\bibinfo {year} {1988})}\BibitemShut
  {NoStop}%
\bibitem [{\citenamefont {Luscher}(2005)}]{Luscher2004}%
  \BibitemOpen
  \bibfield  {author} {\bibinfo {author} {\bibfnamefont {M.}~\bibnamefont
  {Luscher}},\ }\bibfield  {title} {\bibinfo {title} {{Schwarz-preconditioned
  HMC algorithm for two-flavour lattice QCD}},\ }\href
  {https://doi.org/10.1016/j.cpc.2004.10.004} {\bibfield  {journal} {\bibinfo
  {journal} {Comput. Phys. Commun.}\ }\textbf {\bibinfo {volume} {165}},\
  \bibinfo {pages} {199} (\bibinfo {year} {2005})},\ \Eprint
  {https://arxiv.org/abs/hep-lat/0409106} {arXiv:hep-lat/0409106} \BibitemShut
  {NoStop}%
\bibitem [{\citenamefont {Abbott}\ \emph
  {et~al.}(2022{\natexlab{b}})\citenamefont {Abbott} \emph
  {et~al.}}]{Abbott2022b}%
  \BibitemOpen
  \bibfield  {author} {\bibinfo {author} {\bibfnamefont {R.}~\bibnamefont
  {Abbott}} \emph {et~al.},\ }\bibfield  {title} {\bibinfo {title} {{Aspects of
  scaling and scalability for flow-based sampling of lattice QCD}},\
  }\href@noop {} {\  (\bibinfo {year} {2022}{\natexlab{b}})},\ \Eprint
  {https://arxiv.org/abs/2211.07541} {arXiv:2211.07541 [hep-lat]} \BibitemShut
  {NoStop}%
\end{thebibliography}%

\end{document}